\documentclass[aps,prx,twocolumn,superscriptaddress]{revtex4-1}
\usepackage[colorlinks=true,citecolor=blue,linkcolor=blue,urlcolor=blue]{hyperref}
\usepackage{graphicx, nicefrac, textcomp}
\usepackage[normalem]{ulem}
\usepackage{changes}
\usepackage{chngcntr}
\usepackage{amssymb}

\begin{document}

\title{Magnetic correlations in infinite-layer nickelates: an experimental and theoretical multi-method study}
\author{R.~A.~Ortiz$^{1,*}$, P.~Puphal$^{1,*}$, M.~Klett$^{1}$, F.~Hotz$^{2}$, R.~K.~Kremer$^{1}$, H.~Trepka$^{1}$, M.~Hemmida$^{2}$, H.-A.~Krug~von~Nidda$^{2}$, M.~Isobe$^{1}$, R.~Khasanov$^{3}$, H.~Luetkens$^{3}$, P.~Hansmann$^{4}$, B.~Keimer$^{1}$, T.~Sch{\"a}fer$^{1,\dagger}$, and M.~Hepting$^{\dagger,}$}
\affiliation{Max-Planck-Institute for Solid State Research, Heisenbergstra{\ss}e 1, 70569 Stuttgart, Germany\\
$^{2}$Experimental Physics V, Center for Electronic Correlations and Magnetism, University of Augsburg, 86159 Augsburg, Germany\\
$^{3}$Laboratory for Muon Spin Spectroscopy (LMU), Paul Scherrer Institute (PSI), Forschungsstrasse 111, CH-5232 Villigen, Switzerland\\
$^{4}$Department of Physics, University of Erlangen-N{\"u}rnberg, 91058, Erlangen, Germany}

\date{\today}

\begin{abstract}

We report a comprehensive study of magnetic correlations in LaNiO$_{2}$, a parent compound of the recently discovered family of infinite-layer (IL) nickelate superconductors, using multiple experimental and theoretical methods. Our specific heat, muon-spin rotation (\textmu SR), and magnetic susceptibility measurements on polycrystalline LaNiO$_{2}$ show that long-range magnetic order remains absent down to 2 K. Nevertheless, we detect residual entropy in the low-temperature specific heat, which is compatible with a model fit that includes paramagnon excitations. The \textmu SR and low-field static and dynamic magnetic susceptibility measurements indicate the presence of short-range magnetic correlations and glassy spin dynamics, which we attribute to local oxygen non-stoichiometry in the average infinite-layer crystal structure. This glassy behavior can be suppressed in strong external fields, allowing us to extract the intrinsic paramagnetic susceptibility. Remarkably, we find that the intrinsic susceptibility shows non-Curie-Weiss behavior at high temperatures, in analogy to doped cuprates that possess robust non-local spin fluctuations. The distinct temperature dependence of the intrinsic susceptibility of LaNiO$_{2}$ can be theoretically understood by a multi-method study of the single-band Hubbard model in which we apply complementary cutting-edge quantum many-body techniques (dynamical mean-field theory, cellular dynamical mean-field theory and the dynamical vertex approximation) to investigate the influence of both short- and long-ranged correlations. Our results suggest a profound analogy between the magnetic correlations in parent (undoped) IL nickelates and doped cuprates. 
\end{abstract}

\maketitle

\section{Introduction}

Revealing the major mechanism mediating Cooper pairing in unconventional superconductors is one of the defining challenges of condensed matter research \cite{Scalapino2012}. In superconducting cuprates and pnictides, pairing via exchange of magnetic excitations is widely considered to be the most relevant scenario \cite{Scalapino2012, Keimer2015, Si2016, Lee2006}. Yet, the nature of the ``glue" of the superconductivity in Sr or Ca substituted $R$NiO$_2$ ($R = $ La, Pr, Nd) \cite{Li2019,Ariando2020,Osada2020,Li20201,Osada2021, Zeng2021} has not been clarified conclusively \cite{Jiang2020}. Several theoretical \cite{Katukuri2020,Werner2020,Zhou2020,Chang2020,Leonov2020,Wu2020,Sakakibara2020,Kitatani2020,Zhang2020} and first experimental studies \cite{Cui2021,Zhao2021,lu2021} on these infinite-layer (IL) nickelates have suggested the presence of cuprate-like spin fluctuations, which therefore are a prime candidate for providing the attractive interactions underlying their presumably unconventional superconductivity \cite{Wu2020,Wang2021,Botana2021}. Moreover, IL nickelates and cuprates share key similarities  \cite{Anisimov1999}, such as a nominal 3$d^9$ electronic configuration of the Ni$^{1+}$ and Cu$^{2+}$ ions with spin $s = 1/2$. Furthermore, IL nickelates and IL cuprates are isostructural, with Ni (Cu) and O ions arranged in square planar coordination within NiO$_2$ (CuO$_2$) planes that are stacked along the $c$-axis. In case of cuprates --- even without IL structure --- these planes crucially host the superconducting condensate, which emerges upon doping with charge carriers when the parent insulating ground state with long-range antiferromagnetic (AFM) order vanishes.

However, it was realized early that the effective electronic structures of IL nickelates and cuprates are distinct to a certain degree \cite{Lee2004}. Notably, X-ray spectroscopic measurements demonstrated that 3$d$-2$p$ hybridization between Ni and its O ligands is minimal in IL (La,Nd)NiO$_2$ \cite{Hepting2020,Hepting2021}, while strong 3$d$-2$p$ mixing of Cu and O is a hallmark of cuprates. Instead, the presence of states with mixed Ni 3$d$ and rare-earth 5$d$ character was detected \cite{Hepting2020,Hepting2021}, which indicates a distinguished role of the rare-earth spacer layer and was also predicted in \textit{ab initio} calculations \cite{Lee2004, Nomura2019, Sakakibara2020, Botana2020, Wang20202, Been2021}. While it is still under debate whether the hybridized three-dimensional states of weakly interacting 5$d$ electrons play an active role in shaping the low-energy physics of IL nickelates \cite{Zhang20201, Zhang20202, Lechermann2020, Kitatani2020, Wang20202, Karp2020, Karp2021}, recent spectroscopic studies reported that the doped holes in Nd$_{1-x}$Sr$_x$NiO$_2$ reside predominantly in the planar Ni 3$d$ orbitals \cite{Goodge2021}, giving rise to a 3$d^8$ spin singlet state of hole pairs \cite{rossi2020}.

Yet, a coherent picture of the magnetism in IL nickelates has not been established. In fact, previous theoretical approaches often addressed magnetic correlations in terms of static AFM states \cite{Anisimov1999,Lee2004,Kapeghian2020,Zhang20202,Botana2020, Liu2020, Ryee2020, Gu20201, Zhang2021}, while such long-range ordered states have remained elusive in experiments \cite{Crespin1983,Crespin2005,Hayward1999,Hayward2003,Li2019,Kawai2009,Wang20201,Chen2021,lu2021}. This absence is in stark contrast to superconducting cuprates, where the undoped (parent) compounds exhibit commensurate AFM order \cite{Keimer2015}. Notably, strong AFM correlations persist in cuprates in form of spin fluctuations \cite{Birgeneau2006} even when charge carrier doping suppresses long-range order \cite{Keimer2015}. Prominently, the signatures of such fluctuations were detected in doped cuprates in form of paramagnons \cite{LeTacon2011}, which are damped spin excitations that pervade the phase diagram up to highest doping levels \cite{Dean2013,LeTacon2013}. Conversely, it was revealed that similar damped spin excitations emerge in undoped IL nickelates \cite{lu2021}. Thus, it is a pressing issue to clarify the nature of magnetic correlations in IL nickelates with respect to the phenomenology of cuprates in a concerted effort between experiment and theory.  

In addition to the comparison to other unconventional superconductors, IL nickelates are a material class that is intriguing in its own right. For instance, it was recently proposed \cite{Kitatani2020} that IL nickelates could be a genuine realization of the single-band Hubbard model \cite{Hubbard1963, Hubbard1964, Gutzwiller1963, Kanamori1963, Qin2021, Arovas2021}, which is arguably the most paradigmatic model for correlated electron systems. This would allow one to describe long- and short-ranged spatial correlations by means of cutting edge quantum-field theoretical methods. Along these lines the dynamical vertex approximation (D$\Gamma$A) \cite{Toschi2007, Katanin2009}, which is a diagrammatic extension \cite{Rohringer2018} of the dynamical mean-field theory (DMFT) \cite{Metzner1989, Georges1992, Georges1996}, was recently applied to a single-band low-energy representation of Nd-based IL nickelates \cite{Kitatani2020}. The D$\Gamma$A  calculations yielded good agreement with the experimentally observed $T_{\text c}$ of 10 - 15 K in Nd$_{0.8}$Sr$_{0.2}$NiO$_{2}$ thin films \cite{Li2019} which motivates our explicit analysis of the magnetic correlation functions in the same theoretical framework. Additionally, the setting of $s = 1/2$ spins on a square lattice was predicted to bear extraordinary spin correlations \cite{Read1989, Zhitomirsky1996, Samajdar2019} even in absence of long-range order \cite{Anderson1987, Gong2014, Uematsu2018}. Specifically for IL nickelates, proximity to a frustrated quantum critical point was proposed \cite{Choi2020} that could give rise to a conducting spin-liquid phase \cite{Choi2020,Leonov2020}, while other theoretical studies have suggested the occurrence of a spin-freezing crossover due to pronounced Hund's coupling \cite{Werner2020}.  

In this work, we shed new light on the magnetic correlations in polycrystalline LaNiO$_{2}$ using  specific heat ($C_{p}$), muon-spin rotation (\textmu SR), and static ($\chi$) and dynamic ($\chi^\prime$, $\chi^{\prime\prime}$) magnetic susceptibility measurements. Along the lines of previous experiments \cite{Crespin1983,Hayward1999,Crespin2005}, we obtained the IL phase from topotactic reduction of LaNiO$_{3}$ in the perovskite phase. In addition, LaNiO$_{3}$ was employed as a reference for the analysis of $C_{p}$ and the susceptibilities of LaNiO$_{2}$. In the $C_{p}$ of LaNiO$_{2}$ we observe residual entropy at low temperatures, indicating the presence of spin fluctuations and related paramagnon excitations. Our \textmu SR and low-field static and dynamic magnetic susceptibility measurements reveal short-range magnetic correlations and cluster glass-like spin freezing at low temperatures. The glassy dynamics are attributed to subtle non-stoichiometries of the oxygen sublattice, which develop during the synthesis of LaNiO$_2$ via topochemical reduction of LaNiO$_3$. The corrected intrinsic susceptibility $\chi_{\text corr}$ of LaNiO$_{2}$ can be extracted from susceptibility measurements in strong external fields, unveiling a distinct non-Curie-Weiss behavior over a wide temperature range. In particular, we find that $\chi_{\text corr}(T)$ continuously increases with increasing temperature, which is reminiscent of the susceptibility of underdoped cuprates. This temperature dependence of $\chi_{\text corr}$ can be qualitatively understood by a multi-method study of the two-dimensional single-band Hubbard model using two complementary quantum field theoretical methods, dynamical mean-field theory (DMFT), cellular dynamical mean-field theory (CDMFT) and the dynamical vertex approximation (D$\Gamma$A). In the context of this theoretical understanding of the nature of magnetic correlations in IL nickelates, we also comment on the connection between the obtained non-Curie-Weiss behavior and the opening of a pseudogap.

\section{Methods}

\subsubsection{Experimental methods}

Polycrystalline LaNiO$_{3}$ powder was synthesized via the citrate-nitrate method \cite{Alonso1995}. The metal-nitrates Ni(NO$_{3}$)$_{2}$ and La(NO$_{3}$)$_{3}$ and citric acid were dissolved in water. The homogeneous solution was heated on a hotplate to 500 $^\circ$C, which evaporates the water and finally decomposes the nitrates. The oxide was further treated by repeated dry ball-milling and calcination at 
750 $^\circ$C for at least 12 h. Subsequently, the powder was treated in an autoclave with 400 bar O$_{2}$ pressure to assure phase purity and full oxygenation. The autoclave starting temperature of 450 $^\circ$C was decreased to 250 $^\circ$C over 2 days. The LaNiO$_{2}$ IL phase was obtained through topotactic reduction of the  LaNiO$_{3}$ perovskite phase using CaH$_2$ as a reducing agent. Batches of 50 mg LaNiO$_3$ powder were wrapped in aluminum foil with an opening at one end and loaded into quartz tubes with approximately $250$ mg CaH$_2$ powder. To prevent the CaH$_2$ from reacting with air, the procedure was performed in an Ar-filled glove-box and the quartz tubes were sealed under vacuum ($\approx 10^{-7}$ mbar). After sealing, the tubes were heated in a furnace with a ramp rate of $10^{\circ}$C/min and kept at $280^{\circ}$C for $316$ h. Finally, the samples were cooled at a similar rate and the reduced powder was then extracted. During the entire process, the nickelate and CaH$_2$ powder have not been in direct contact. 

LaNiO$_{3}$ and the reduced phase were characterized by powder X-ray diffraction (PXRD) (see App.~\ref{app:structure}). The refined structural parameters [Table \ref{refinement}] are in good agreement with previous reports \cite{GarciaMunoz1992, Hayward1999,Crespin2005}. Furthermore, the refinement of LaNiO$_{3}$ [Fig.~\ref{FigA1}a] indicates that the citrate-nitrate method and subsequent autoclave treatment yields highly pure powders without traceable secondary phases. Also the data on the reduced sample can be refined assuming a single phase of IL LaNiO$_{2}$ after a reduction time of 316 h [Fig.~\ref{FigA1}b], without indications for impurities due to incomplete reduction or decomposition, such as LaNiO$_{2.5}$, NiO/Ni, or La$_2$O$_3$. For shorter reduction times, residues of phases with excess oxygen were detected, for instance we found that after 196 h the reduced sample contained $14.7$\,wt\% of LaNiO$_{2.5}$. The phase purity was further confirmed on pressed LaNiO$_{3}$ and LaNiO$_{2}$ pellets by scanning electron microscopy with energy dispersive x-ray spectroscopy (EDS) and no agglomerations of elemental Ni were found (see App.~\ref{app:structure}). Inductively coupled plasma mass spectrometry (ICP-MS) indicated ideal stoichiometries (within the experimental error) of La$_{0.99(1)}$Ni$_{0.99(1)}$O$_{2.99(3)}$ and La$_{0.99(1)}$Ni$_{0.99(1)}$O$_{2.02(3)}$, respectively (see App.~\ref{app:structure}).

The specific-heat data were collected with the standard options of a Physical Property Measurement System (PPMS, Quantum Design) using a thermal relaxation method on cold pressed pellets of LaNiO$_{3}$ and LaNiO$_{2}$ powder. The low-temperature dc susceptibility measurements were performed using a vibrating sample magnetometer (MPMS VSM SQUID, Quantum Design) and the ac susceptibility measurements using a Physical Property Measurement System (PPMS, Quantum Design) with an ACMS option. High-temperature magnetization measurements were carried out in a Quantum Design MPMS3 setup.

The \textmu SR experiments were performed on the General Purpose Surface-Muon Instrument (GPS) \cite{GPS} at the Paul Scherrer Institute (PSI) in Switzerland. The GPS instrument was operated in veto mode, which minimizes the background signal below the detection limit. The data analysis was performed with the free software package musrfit \cite{SUTER201269}.

\subsubsection{Theoretical methods}
\label{sec:theory}

For the theoretical modeling of LaNiO$_{2}$, we employ the parameters given in a recent single-band Hubbard model study for NdNiO$_{2}$ \cite{Kitatani2020}: We set the onsite Coulomb interaction to $U\!=\!3.2$ eV, and the tight-binding parameters to  $t\!=\!395$ meV, $t'\!=\!-95$ meV and $t''\!=\!47$ meV  (expressed via the nearest neighbor transition amplitude $t$, these parameters read $U\!=\!8t$, $t'\!=\!-0.25t$ and $t''\!=\!0.12t$). Like for NdNiO$_{2}$, we neglect a (small) hopping integral in $c$-direction also for LaNiO$_{2}$. Note that the electronic structure of LaNiO$_{2}$ corresponds more closely to a half-filled $d_{x^2-y^2}$ band than NdNiO$_{2}$ \cite{Kitatani2020}, suggesting that the half-filled model is even more appropriate for the former case. Hence we use a half-filled model for all calculations shown. 

In order to obtain the magnetic susceptibility we apply multiple many-body techniques for correlated systems to the same model. Such multi-method studies have proven highly valuable \cite{LeBlanc2015, Schaefer2021, Wietek2021} for the calculation of observables in challenging parameter regimes of the model. Here we employ the dynamical mean-field theory (DMFT, \cite{Metzner1989,Georges1992,Georges1996}) and two complementary extensions thereof: (i) the dynamical vertex approximation (D$\Gamma$A, \cite{Toschi2007, Katanin2009}), which is a diagrammatic extension \cite{Rohringer2018} of DMFT and the cellular dynamical-mean field theory (CDMFT), a real-space cluster extension \cite{Maier2005} of DMFT. We apply the D$\Gamma$A in its ladder version with Moriyaesque $\lambda$-corrections in the spin channel \cite{Rohringer2018b} and the CDMFT on a $N_c\!=\!4\!\times\!4\!=\!16$ site cluster \cite{Maier2005,Klett2020}.

DMFT includes all local quantum fluctuations of the system, however, entirely neglects spatial correlations. CDMFT captures non-local correlations exactly up to the cluster size. Cluster extensions of DMFT exhibit a non-Curie-Weiss behavior in the susceptibility \cite{Huscroft01,Macridin06,Musshoff2021}, associated with the onset of the pseudogap regime of cuprates \cite{Chen17,Werner2020a}. These methods are controlled and unbiased w.r.t. fluctuation channels. Let us note that a non-Curie-Weiss behavior was also recently observed in a minimally-entangled typical thermal states study on finite cylinders \cite{Wietek2021b} as well as in a diagrammatic Monte Carlo study in the weak-coupling regime of the Hubbard model \cite{Kim2020}.

Our complementary method, the D$\Gamma$A, is able to include short- and long-range spatial correlations of the chosen (magnetic) channel on equal footing. The D$\Gamma$A has proven particularly successful in the vicinity of second order phase transitions \cite{Rohringer2011,Schaefer2017,Schaefer2019}, for the estimation of superconducting transition temperatures \cite{Kitatani2019,Kitatani2020}, for obtaining insights into the physics of layered materials and dimensional crossovers \cite{Schaefer2015c,Klebel2021} and in comparison to numerically exact benchmark calculations in the weak-coupling regime of the Hubbard model \cite{Schaefer2021}. The (isolated)  finite-size systems have been diagonalized exactly with PYED in the TRIQS \cite{TRIQS} package. For further calculational details see App.~\ref{app:comp}.

\section{Results}

\subsection{Specific heat}

Nickelates in the IL phase can be obtained via the topotactic removal of oxygen from a precursor $R$NiO$_{3}$ ($R$ = La, Pr, or Nd) perovskite phase \cite{Crespin1983,Crespin2005,Hayward1999,Hayward2003,Li2019,Kawai2009,Wang20201}. In this work, we reduce highly pure LaNiO$_3$ powder to LaNiO$_{2}$ (see Methods). Possible admixture of secondary phases lies below the detection threshold of our PXRD characterization [see App.~\ref{app:structure}]. 

\begin{figure}[tb]
\includegraphics[width=1.0\columnwidth]{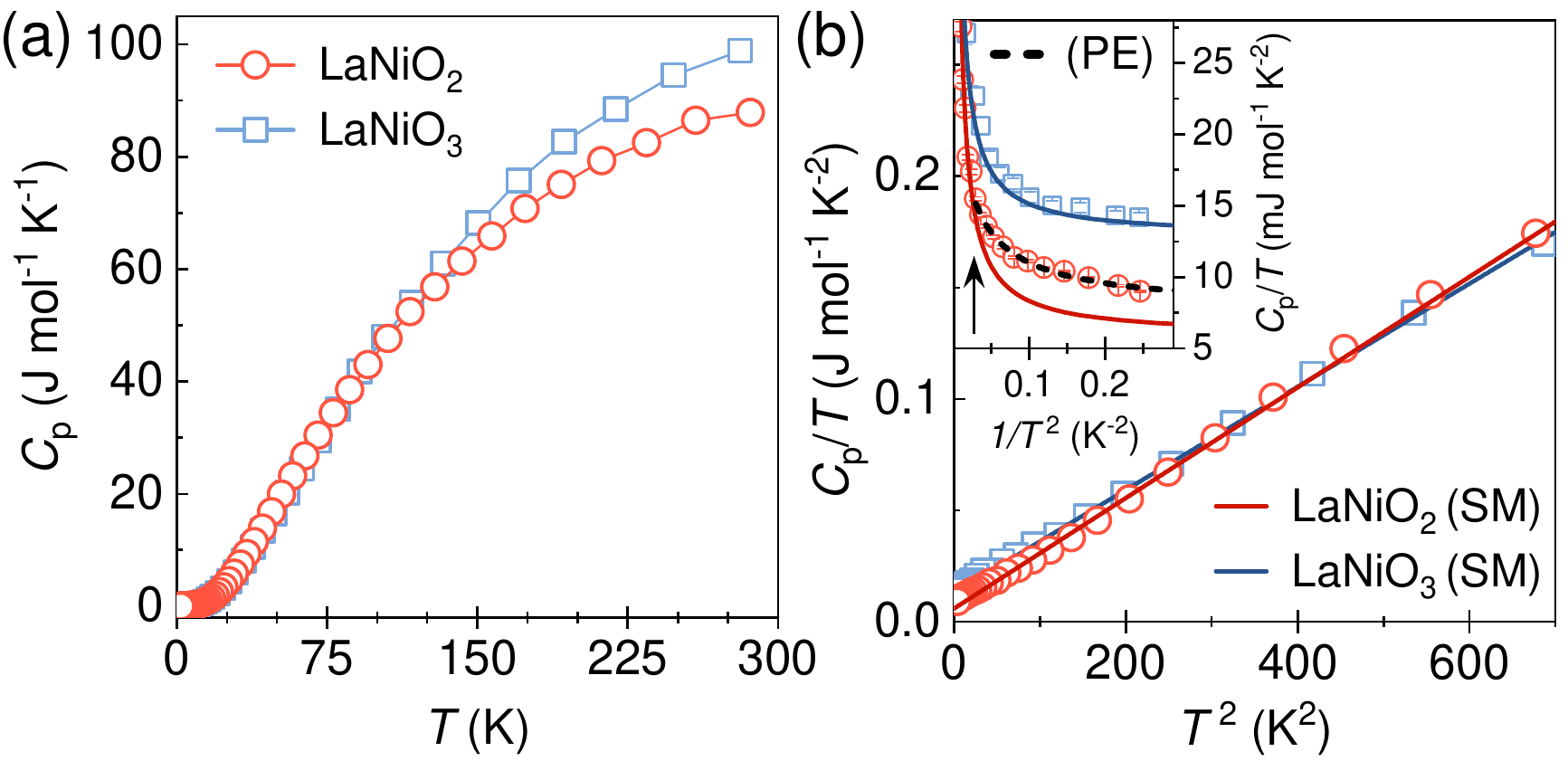}
\caption{(a) Specific heat $C_{p}$ of polycrystalline LaNiO$_{2}$ (red) and LaNiO$_3$ (blue). (b) $C_{p}/T$ as a function of $T^2$. Solid lines are standard model (SM) fits of LaNiO$_{2}$ and LaNiO$_3$ between $T = 6$ and $25$\,K according to $C_{p} =\gamma T +\beta T^3$ (see text). The inset shows $C_{p}/T$ as a function of $1/T^2$. The black arrow centered at $T = 6$\,K indicates the onset of a deviation from the extrapolated SM fit (solid red line) for LaNiO$_{2}$. The dashed black line is a  model fit of LaNiO$_{2}$ between $T = 2$ and $6$\,K according to $C_{p} =\gamma T +\beta T^3 + \alpha T^3 \ln(T)$, which accounts for an extra contribution due to paramagnon excitations (PE).}
\label{Cpheat}
\end{figure}

\begin{figure}[tb]
\includegraphics[width=1.0\columnwidth]{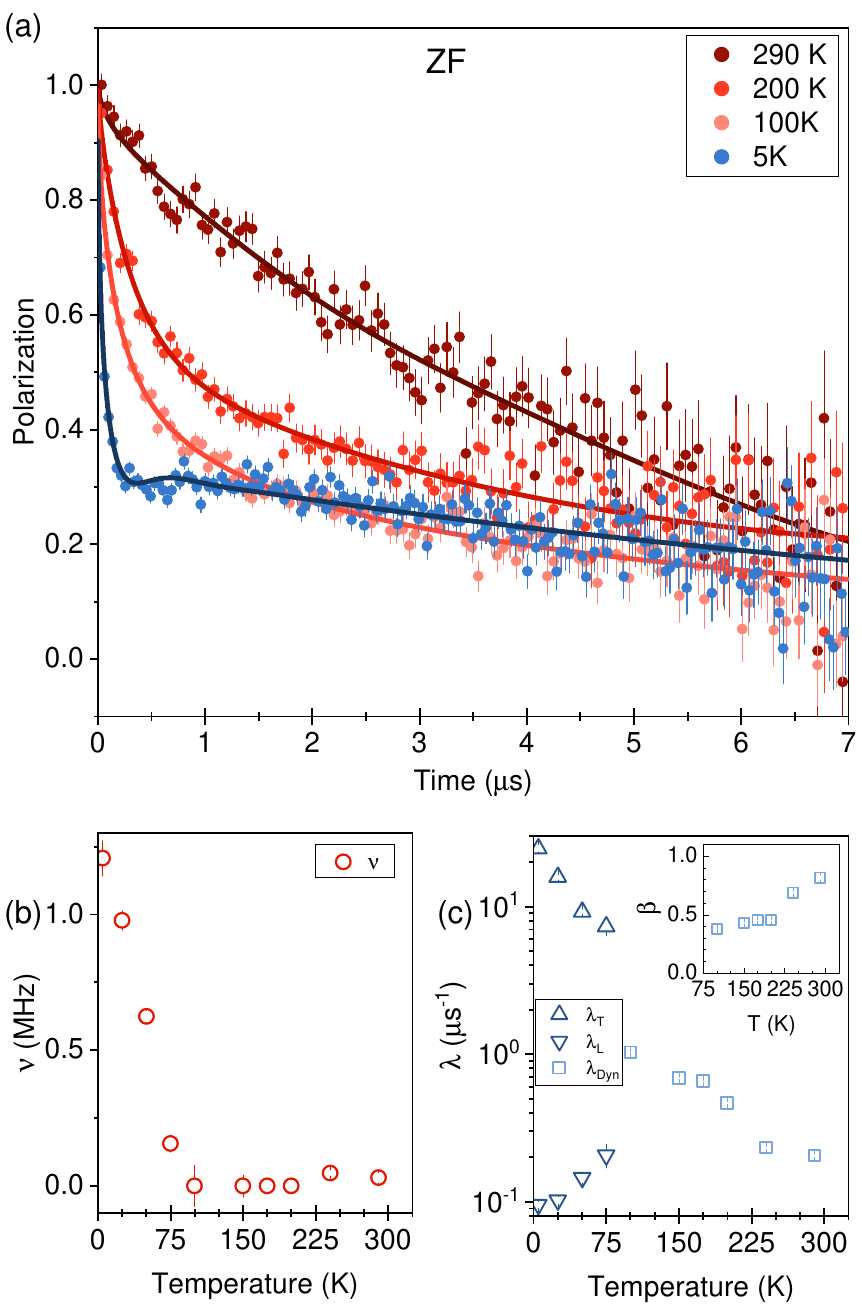}
\caption{(a) Zero-field (ZF) \textmu SR spectra of LaNiO$_2$ at representative temperatures. Solid black lines are fits of the polarization $P(t)$ (see text). (b) Temperature dependence of the  parameter $\nu$, which is indicative of the internal magnetic field. (d) Temperature dependence of the parameters $\lambda_{\text{L}}$, $\lambda_{\text{T}}$, and  $\lambda_{\text{Dyn}}$, extracted from fits to regimes below 75\,K and above 100\,K, respectively.  The inset shows the evolution of the parameter $\beta$ for temperatures above 100\,K obtained from a simple stretched exponential fit. 
}
\label{musr}
\end{figure}

\begin{figure*}[tb]
\includegraphics[width=2.0\columnwidth]{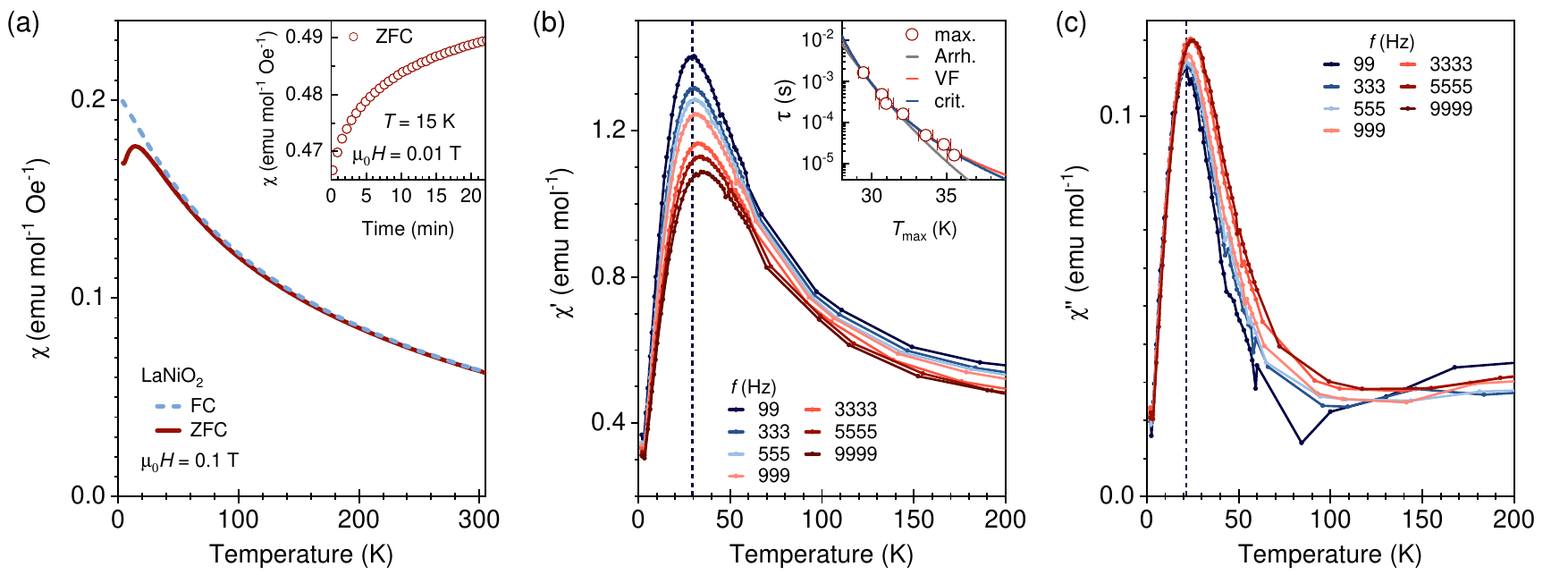}
\caption{Magnetic susceptibility of LaNiO$_{2}$ in small external fields. (a) Static susceptibility $\chi$ measured upon heating after zero-field cooling (ZFC, solid line) and field-cooling (FC, dashed line), respectively, in an external field of $\mu_0 H = 0.1$\,T. The inset shows the evolution of $\chi$ (ZFC, $\mu_0 H = 0.01$\,T) as a function of time for an intermittent stop at $T = 15$\,K. (b) Real part of the dynamic susceptibility $\chi^\prime$ for different ac drive frequencies $f$. The static external field is zero and the maximum field amplitude of the ac field is  $4 \cdot 10^{-4}$\,T. Vertical dashed lines indicate the positions of the maximum for $f = 99$\,Hz. The inset shows the relation between the temperature $T$ of the maximum of $\chi^\prime$ and the time period $\tau = \left(2\pi f \right)^{-1}$ of the ac field. The solid gray, red, and blue lines are fits with an Arrhenius (Arrh.), a Vogel-Fulcher (VF), and a critical dynamical scaling (crit.) law, respectively. (c) Imaginary part of the dynamic susceptibility $\chi^{\prime\prime}$.}
\label{smallfields}
\end{figure*}

As a first step to explore magnetic correlations in LaNiO$_{2}$ we determine the temperature dependence of the specific heat ($C_{p}$) between $2$ and $297$\,K [Fig.~\ref{Cpheat}]. Characteristic anomalies in $C_{p}$ not only reveal the presence of phase transitions, but can also provide valuable information in absence of long-range order, for instance about spin fluctuations \cite{Doniach1966,Rasul1988,Woelfle1990} or frustrated spin states \cite{Cheng2011}. As expected, $\lambda$-like anomalies that correspond to magnetic and/or structural phase transitions are absent in $C_{p}(T)$ of LaNiO$_{2}$ [Fig.~\ref{Cpheat}a]. This is consistent with neutron diffraction experiments on polycrystalline LaNiO$_{2}$ and NdNiO$_{2}$ that reported a lack of magnetic Bragg reflections \cite{Hayward1999,Hayward2003}. A qualitative understanding of $C_{p}$ of LaNiO$_{2}$ can be gained from a direct comparison to the precursor phase LaNiO$_3$, whose $C_{p}$ is remarkably similar at first glance [Fig.~\ref{Cpheat}a], in spite of  different crystallographic symmetries [Tab. \ref{refinement}]. In particular, below $120$\,K the specific heat of the two compounds is almost indistinguishable, which is likely related to the similarity of their La-Ni sublattice [Tab. \ref{refinement}], which gives the major contribution to $C_{p}$ in this temperature range. In contrast, phonons involving the oxygen sublattice --- which is substantially altered upon reduction --- mostly play a role at higher temperatures and can be responsible for the slightly larger $C_{p}$ of LaNiO$_{3}$ above $120$\,K [Fig.~\ref{Cpheat}a].

Figure~\ref{Cpheat}b shows the specific heat below $25$\,K, which in case of LaNiO$_3$ can be described by $C_{p}=\gamma T +\beta T^3$ \cite{Sanchez1993}, according to the standard model for a non-magnetic solid. A fit between 6 and 25\,K results in a Sommerfeld coefficient $\gamma=12.2(4)$\,mJ\,$\cdot$\,mol$^{-1}\,\cdot$\,K$^{-2}$, a lattice contribution $\beta=0.234(8)$\,mJ\,$\cdot$\,mol$^{-1}\,\cdot$\,K$^{-4}$, and a Debye temperature of $\Theta_{D}=346$\,K, in good agreement with previous studies on polycrystalline LaNiO$_3$ \cite{Sanchez1993}. For LaNiO$_{2}$, the same fit  yields $\gamma= 4.4(6)$\,mJ\,$\cdot$\,mol$^{-1}\,\cdot$\,K$^{-2}$, $\beta=0.255(6)$\,mJ\,$\cdot$\,mol$^{-1}\,\cdot$\,K$^{-4}$, and $\Theta_{D}=336$\,K. Thus, the lattice contribution $\beta$ remains almost unchanged in comparison to LaNiO$_3$, whereas the electronic contribution $\gamma$ of LaNiO$_2$ decreases significantly. In general, such decrease can be related to a less correlated electronic character and/or reduced metallicity, which is still an open question for LaNiO$_2$: Electrical transport measurements on high-quality epitaxial LaNiO$_2$ thin films revealed a metallic in-plane conductivity \cite{Ikeda2016}, whereas powder measurements possibly suffer from extrinsic effects, such as inferior contacts between grain boundaries, and the anisotropy between in-plane and $c$-axis transport, rationalizing the observed non-metallic behavior \cite{Hayward1999}. 

The inset in Fig.~\ref{Cpheat}b focuses on the evolution of the specific heat towards the lowest measured temperatures and shows $C_{p}/T$ as a function of $1/T^2$. As expected for the non-magnetic Fermi-liquid LaNiO$_{3}$ \cite{Eguchi2009}, we find that the data below $6$\,K are well described by an extrapolation of the standard model fit $C_{p}=\gamma T +\beta T^3$ performed between $25$ and $6$\,K. However, a similar extrapolation of a standard model fit of LaNiO$_{2}$ does not capture the measured data below $6$\,K [inset in Fig.~\ref{Cpheat}b]. This deviation is small in absolute terms, but nevertheless statistically significant. While we cannot exclude that this residual entropy emerges due to disorder in the polycrystalline powder, we emphasize that such an additional contribution to $C_{p}$ can typically arise from the emission and reabsorption of persistent spin fluctuations or damped spin waves (paramagnons) \cite{Doniach1966,Rasul1988,Woelfle1990}. Employing a large $U$ Hubbard model with parameter choices corresponding to the paramagnetic, metallic regime at half filling, the spin fluctuation term for the specific heat follows as $T^3 \ln(T)$ \cite{Woelfle1990}. Thus, we fit the specific heat of LaNiO$_{2}$ between $2$ and $6$\,K by $C_{p}=\gamma T +\beta T^3 + \alpha T^3 \ln(T)$ and obtain $\gamma= 7.7(9)$ mJ\,$\cdot$\,mol$^{-1}\,\cdot$K$^{-2}$, $\beta=0.5(1)$\,mJ\,$\cdot$\,mol$^{-1}\,\cdot$\,K$^{-4}$, and $\alpha= -0.16(1)$\,mJ\,$\cdot$\,mol$^{-1}\,\cdot$\,K$^{-2}$. Since such fit provides an adequate description of the data below $6$\,K, it can be an indication of the presence of paramagnons in LaNiO$_{2}$. Notably, paramagnons are not only a hallmark of doped cuprates \cite{LeTacon2011, Dean2013}, but damped spin excitations in absence of long-range order were recently also detected with RIXS in thin films of the IL nickelate NdNiO$_{2}$ \cite{lu2021} and the structurally related trilayer nickelate Pr$_4$Ni$_3$O$_8$ \cite{Lin2021}.

\begin{figure*}[tb]
\includegraphics[width=2.0\columnwidth]{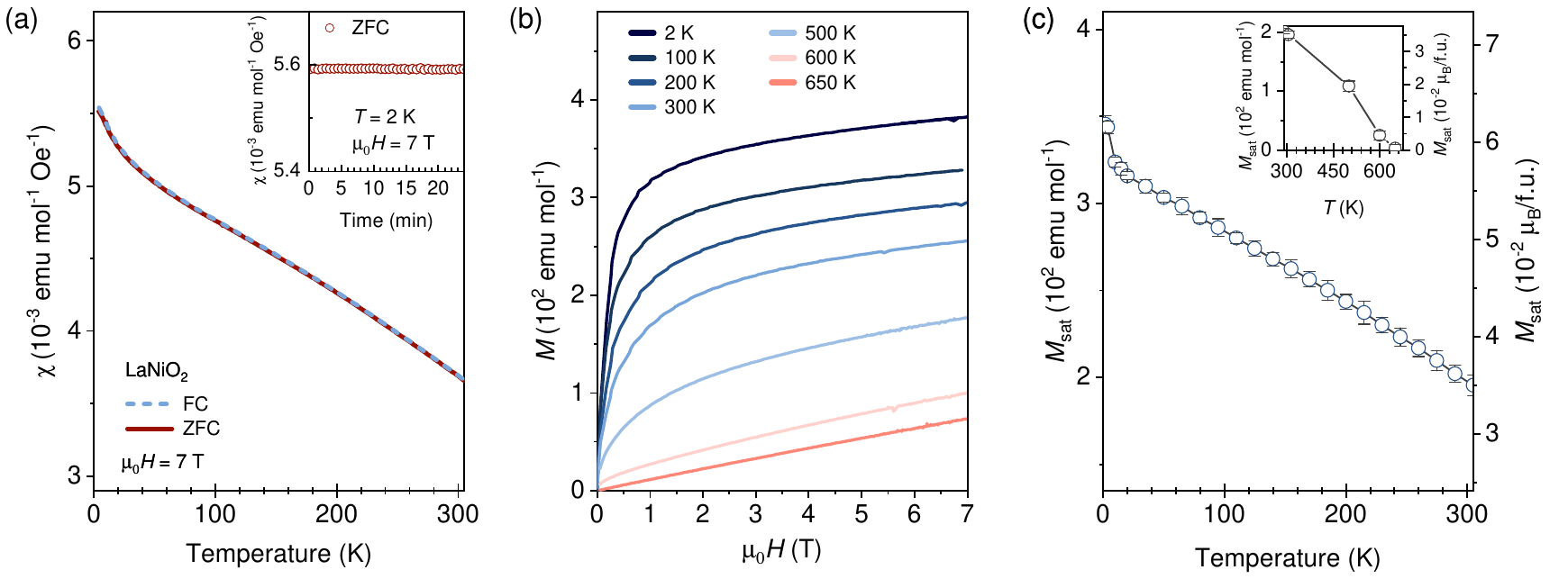}
\caption{Magnetic susceptibility of LaNiO$_{2}$ in strong external fields. (a) ZFC and FC susceptibility $\chi$ in an external field $\mu_0 H = 7$ T. The inset shows the evolution of $\chi$ (ZFC) as a function of time for an intermittent stop at $T = 2$\,K. (b) Isothermal magnetization between 0 and 7 T at representative temperatures. (c) Saturation magnetization $M_{\text{sat}}$ extracted from linear fits of the isothermal magnetization curves between 5 and 7 T for temperatures between 2 and $305$\,K (see App.~\ref{app:suscept}). The inset shows $M_{\text{sat}}$ extracted from isotherms measured at higher temperatures.  
}
\label{strongfields}
\end{figure*}

\subsection{Muon-spin rotation}

Insights into quasi-static and dynamical magnetic processes can be gained from \textmu SR, with positive muons stopping on interstitial lattice sites and acting as a sensitive local probe for small internal magnetic fields and ordered magnetic volume fractions in a sample. Figure~\ref{musr}a displays representative zero-field (ZF) \textmu SR spectra of LaNiO$_{2}$ for temperatures between 290 and 5\,K. The full set of ZF \textmu SR spectra is shown in Fig.~\ref{FigU2}. In accordance with the absence of a magnetic transition in our $C_{p}$ measurements, the \textmu SR spectra do not show well-defined oscillations that would be indicative of long-range magnetic order, as observed for instance in the AFM trilayer nickelate La$_4$Ni$_3$O$_8$ \cite{Bernal2019}. Nonetheless, complementary measurements in longitudinal fields (LF) at 295 K reveal that fluctuating moments are present in the sample [Fig.~\ref{FigU1}]. The ZF spectra below 75\,K [Fig.~\ref{musr}a and Fig.~\ref{FigU2}] exhibit a strongly damped oscillation with a first local minimum around 0.25\,\textmu s, suggesting the occurrence of pronounced short-range order at these temperatures. Moreover, the long-time tail of the 5 K spectrum lies above the 100 K spectrum and approaches a value of 1/3 of the initial polarization, which signals quasi-static magnetism as observed for instance below the spin freezing transition in spin glasses \cite{PhysRevB.31.546}. Figure~\ref{musr}b shows the ZF \textmu SR frequency $\nu(T)$ from fitting the ZF data as described below, suggesting a crossover with an onset between 75 and 100\,K, which we attribute to the onset of glassy behavior, where magnetic moments freeze at low temperatures with random orientations while long-range ferromagnetic or AFM order remain absent \cite{Binder1986}. We consider two fitting regimes for the data, $i.e.$ purely dynamic behavior $P(t) = P_0 e^{-(\lambda_{Dyn} t)^\beta}$ above 75 K, and slowly fluctuating/quasi-static behavior which is described by $P(t) = 1/3 e^{-\lambda_{\text{L}} t} + 2/3 \cos(2\pi \nu t) e^{-(\lambda_{\text{T}} t)^\beta}$ [Fig.~\ref{musr}c] below this temperature. Here, $\lambda_{\text{T}}$ and $\lambda_{\text{L}}$ are the transverse and longitudinal relaxation rates. The 2/3 (transverse) and 1/3 (longitudinal) components reflect the polycrystalline nature of the sample leading to a powder average of the internal fields with respect to the initial muon spin direction.

Figure~\ref{musr}c and the inset in Fig.~\ref{musr}c show the temperature evolution of the muon spin relaxation rate $\lambda$ and the stretch parameter $\beta$ in the dynamic regime between 295 and 75 K, respectively. The observed increase of $\lambda$ with decreasing temperature and variation of $\beta$ from approximately 1 to 0.5 indicating a broad distribution of magnetic relaxation times typical for a spin glass prior to the freezing transition  \cite{PhysRevB.31.546, PhysRevB.64.054403, PhysRevLett.72.1291}. Notably, the temperature range across which this variation occurs is relatively broad. This is a typical feature of heterogeneous samples, with the muons experiencing both, strong and small fields in the sample. Thus, polycrystalline LaNiO$_2$ likely exhibits a spatially inhomogeneous distribution of magnetic correlations, possibly due to the presence of magnetic clusters or islands. Fluctuations of spatially separated correlated clusters are typically slow, which is compatible with the extracted time scales in Fig.~\ref{musr}c. Below 75 K the dynamic relaxation rate $\lambda_{L}$ decreases consistent with a continuous slowing-down also in the quasi-static regime. The observed peak in the dynamic relaxation rate is a fingerprint of magnetic fluctuations slowing down and crossing the time window of the measurement. The transverse relaxation rate $\lambda_{T}$ and the frequency $\nu$ of the overdamped oscillation increase below 75 K indicative of the growing magnetic correlations. 

In summary, the \textmu SR data suggest the presence of glassy magnetic behavior with an inhomogeneous spatial and temporal distribution. Nevertheless, especially the observation of a ZF \textmu SR oscillation with a large relaxation rate $\lambda_{T}$ of up to 25  $\mu s^{-1}$  together with the typical separation of the spectra in a 1/3 and a 2/3 component is a strong indication that the  magnetic behavior originates from the bulk of the sample.

\subsection{Magnetic susceptibility}

As a next step, we investigate the static magnetic susceptibility ($\chi$) of LaNiO$_{2}$. Figure~\ref{smallfields}a shows $\chi(T)$ of LaNiO$_{2}$ measured in a relatively small external field of $0.1$\,T. In accordance with previous experiments \cite{Crespin1983,Crespin2005,Hayward1999,Hayward2003,Wang20201} and our $C_p$ and \textmu SR results, the magnetization signal lacks any signatures of long-range magnetic ordering between $4$ and $300$\,K. However, we observe several salient features, which were not reported in early studies of polycrystalline LaNiO$_{2}$ \cite{Crespin1983,Hayward1999}. In particular, we detect an irreversibility between the zero-field cooled (ZFC) and field-cooled (FC) magnetization branches at low temperatures and a cusp-like feature of the ZFC branch centered around 17\,K. Such behavior of the ZFC and FC magnetization is characteristic for spin glasses. Furthermore, hysteretic behavior is observed in measurements of the isothermal magnetization  [Fig.~\ref{FigB1}b], which is enhanced at temperatures corresponding to the region of the cusp of the ZFC curve and below. Closely similar ZFC-FC splittings and magnetic hysteresis were also reported in recent studies on polycrystalline $R$NiO$_2$ and Pr$_4$Ni$_3$O$_8$ powders \cite{lin202101, Huangfu2020}, as well as La$_{1-x}$Ca$_{x}$NiO$_{2+\delta}$ crystals \cite{Puphal2021}. In principle, the maximum of the cusp can be associated with the freezing temperature $T_{\text{f}}$ of the glass, however, note that our measurement of $\chi(T)$ does not probe the equilibrium state of the spins. Specifically, the cusp feature of the ZFC curve shifts to higher temperatures when the $\chi(T)$ measurements are conducted in smaller external static fields [Fig.~\ref{FigB1}a]. Moreover, $\chi$ evolves as a function of the measurement time [inset in Fig.~\ref{smallfields}a], which is an additional indication that dynamical processes are at play. Thus, the observed maximum of the cusp at 17\,K in Fig.~\ref{smallfields}a is only an approximate indicator of $T_{\text{f}}$.

Detailed insights into the dynamics and the nature of the spin glass phase can be gained from measurements of the real and imaginary parts of the susceptibility $\chi^\prime$ and $\chi^{\prime\prime}$ in an ac drive field. Figures~\ref{smallfields}b,c, show that the ac susceptibilities of LaNiO$_{2}$ exhibit a pronounced cusp/peak at low temperatures, which is reminiscent of the cusp in the dc susceptibility [Fig.~\ref{smallfields}a and Fig.~\ref{FigB1}a]. We note that in particular the onset of the peak in $\chi^{\prime\prime}(T)$ [Fig.~\ref{smallfields}c] coincides with the onset of the spin freezing identified in the \textmu SR measurements [Fig.~\ref{musr}]. With increasing drive frequency $f$, the maximum of the peak in $\chi^\prime(T)$ shifts to higher temperatures and its amplitude decreases. Similarly, the position of the peak in $\chi^{\prime\prime}(T)$ increases, whereas its amplitude and width also increase with increasing $f$ [Fig.~\ref{smallfields}c]. Such behavior of $\chi^\prime$ and $\chi^{\prime\prime}$ is typical for both, canonical and cluster spin glasses \cite{Binder1986}, while the \textmu SR results strongly indicated spatial inhomogeneities, \textit{ i.e.} our sample consists of an assembly of interacting magnetic clusters. 

The interaction strength between magnetic clusters can be assessed from the scaling behavior \cite{Binder1986} of the frequency dependent peak position of $\chi^\prime (T)$ [Fig.~\ref{smallfields}b]. To this end, the frequency dependent peak position in $\chi^\prime$ is fitted using different laws  [inset in Fig.~\ref{smallfields}b], which are appropriate for increasing interaction strengths, including an Arrhenius, a Vogel-Fulcher (VF), and a critical slowing down law. An Arrhenius law can capture systems of low-dimensional non-interacting magnetic entities and is expressed as $\tau = \tau_0 \exp(E_{\text{a}}/k_{\text{B}}T)$, where $\tau = \left(2\pi f \right)^{-1}$ is the relaxation time, $\tau_0$ a characteristic time, $E_{\text{a}}$ an energy barrier separating two low energy states, and $k_{\text{B}}$ the Boltzmann constant. Here, we obtain $\tau_0 = 4.6 \cdot 10^{-17}$ s and $E_{\text{a}}/k_{\text{B}} = 918$ K. The nonphysically small value of $\tau_0$ and the extremely large $E_{\text{a}}$ \cite{Binder1986} suggest that LaNiO$_{2}$ does not fall into the non-interacting Arrhenius limit. In addition, we can rule out superparamagnetic behavior, which exhibits typical values of $\tau_0$ between $10^{-9}$ and $10^{-13}$ s \cite{Binder1986,ichiyanagi1996}. In presence of moderate inter-cluster couplings in a spin glass, a VF law according to $\tau = \tau_0 \exp(E_{\text{a}}/k_{\text{B}}(T - T_{\text{0}}))$ can be employed, where $T_{\text{0}}$ is the glass temperature, which  typically is similar to the freezing temperature. For LaNiO$_{2}$, we obtain $\tau_0 = 1.2 \cdot 10^{-8}$ s, $E_{\text{a}}/k_{\text{B}} = 118.4 $ K, and $T_{\text{0}} = 19.4 $ K. This yields $E_{\text{a}}/(k_{\text{B}}\tau_0) \approx 6$, indicating that LaNiO$_{2}$ is located in the weak to intermediate coupling regime, whereas $E_{\text{a}}/(k_{\text{B}}\tau_0) \ll 1$ would correspond strong coupling. The obtained time scale $\tau_0$ is comparable with the relaxation rate $\lambda_{\text{L}}$ extracted from the \textmu SR measurements, suggesting that the same freezing phenomenon is probed by the two complementary techniques. For completeness, we also test a strongly interacting critical dynamical scaling law according to a power law of the form $\tau = \tau_0 ((T - T_{\text{f}})/T_{\text{f}})^{-z\nu}$, with the freezing temperature $T_{\text{f}}$ in the limit $f \rightarrow 0$ and the dynamic critical exponent $z\nu$ that accounts for a critical slowing down of the dynamics of the magnetic entities in proximity to $T_{\text{f}}$. A fit of the data in the inset in Fig.~\ref{smallfields}b results in $\tau_0 = 1.6 \cdot 10^{-6}$ s, $T_{\text{f}} = 19.5$ K, and $z\nu = 10.8$. While the critical exponent adopts an acceptable value, $\tau_0$ is larger than that of typical canonical and cluster glasses \cite{Binder1986}. In consequence, our polycrystalline LaNiO$_{2}$ sample is best described by a VF law with weak to intermediate coupling between magnetic clusters. We note that a VF law was also employed to fit the susceptibility of polycrystalline Pr$_4$Ni$_3$O$_8$ \cite{Huangfu2020} and $R$NiO$_2$ \cite{lin202101}, although a stronger coupling $E_{\text{a}}/(k_{\text{B}}\tau_0)$ was reported.

The observation of glassy behavior with weak to intermediate coupling strength in the \textmu SR and dynamic susceptibility measurements raises the question whether glassiness is an intrinsic property of LaNiO$_2$, or arises from possible impurities/secondary phases. In the following, we address this issue by susceptibility measurements in strong external fields [Fig.~\ref{strongfields}]. In the case of ferromagnetic impurities in reduced powder samples \cite{Jin2018,Jin2020}, it is expected that their magnetization signal saturates in sufficiently strong fields, yielding only a constant contribution after the field exceeds a certain threshold value \cite{Hayward1999}. Moreover, large fields suppress the glassy dynamics [Fig.~\ref{FigB1}a] and underlying intrinsic magnetic correlations can be exposed. Figure~\ref{strongfields}a displays the static susceptibility $\chi$ of our polycrystalline LaNiO$_{2}$ sample in a field of $7$\,T. In contrast to the low field measurement [Fig.~\ref{smallfields}a], an irreversibility between the ZFC and FC curves is not present, thus confirming that spin glass effects do not dominate the signal at this field strength. Furthermore, $\chi$ does not depend on measurement time or history [inset of Fig.~\ref{strongfields}a]. Overall, $\chi$ increases almost linearly with decreasing temperature, while a subtle upturn at lowest temperatures and a broad hump centered around $200$\,K can be identified. To determine whether these features are related to the intrinsic paramagnetic susceptibility of LaNiO$_{2}$, or induced by (ferromagnetic) impurities, we measured magnetization-field isotherms $M(H)$ at various temperatures [Fig.~\ref{strongfields}b]. The isotherms show a pronounced non-linearity for small applied fields, which decreases for isotherms measured at higher temperatures and vanishes between 600 and 650 K. Since the Curie temperature of Ni is $\sim 620$ K, the non-linearity is likely due to the presence of a small amount of Ni impurities, which we estimate to be less than 2.5 \,wt\% (see App.~\ref{app:structure} and App.~\ref{app:suscept}). The saturation magnetization $M_{\text{sat}}$ can be determined from linear fits to the isotherms at high fields in the range between $5$ and $7$\,T (see App.~\ref{app:suscept}). The temperature dependence of $M_{\text{sat}}$ [Fig.~\ref{strongfields}c] qualitatively resembles $\chi(T)$ in Fig.~\ref{strongfields}a, suggesting the temperature dependence of $\chi(T)$ in Fig.~\ref{strongfields}a is dominated by the signal of the ferromagnetic Ni impurities and does not reflect the intrinsic susceptibility of LaNiO$_{2}$. 

\begin{figure}[tb]
\includegraphics[width=.9\columnwidth]{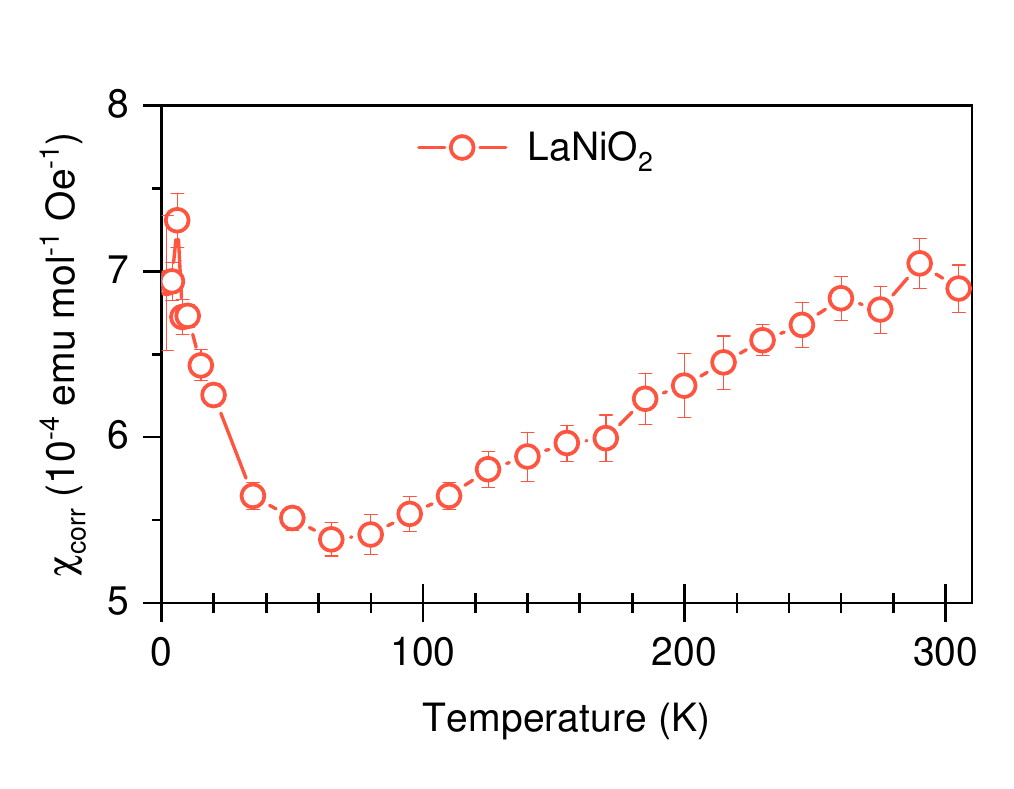}
\caption{Corrected paramagnetic susceptibility $\chi_{\text corr}$ extracted from isothermal magnetization curves by the Honda-Owen method (see App.~\ref{app:suscept}).}
\label{Chi_corr}
\end{figure}

\begin{figure*}[t!]
\centering
\includegraphics[width=1.0\textwidth]{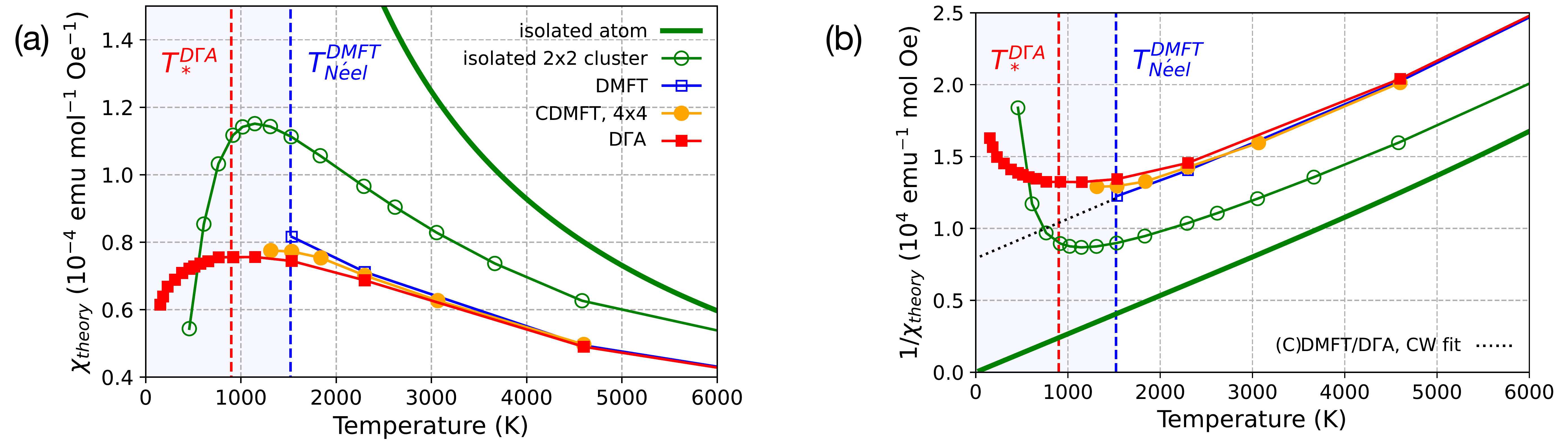}
\caption{\label{fig:theory} (a) dc-susceptibility $\chi_{\text{theory}}$ and (b) its inverse  calculated for a two-dimensional single-band Hubbard model of LaNiO$_{2}$ by various many-body techniques. CDMFT (filled orange circles) and D$\Gamma$A (filles red squares) exhibit a deviation from Curie-Weiss behavior at low $T$ when significant non-local correlations set in [the dotted black line in (b) shows a respective Curie-Weiss fit from the high-$T$ regime of all three techniques].  This regime is already signalled in DMFT as an (antiferromagnetically) ordered phase for $T\!<\!T_{\text{N{\'e}el}}^{\text{DMFT}}$ (blue shaded region). For comparison to finite-size systems also the atomic values (solid green line, Curie law) and the ones of an isolated $2\!\times\!2$ cluster (open green circles) are shown.
}
\end{figure*}

Nevertheless, a distinction between the ferromagnetic Ni background and intrinsic paramagnetic behavior of the majority phase is possible by applying the Honda-Owen method \cite{Honda1910,Owen1912} to the isotherms measured between $5$ and $7$\,T (see App.~\ref{app:suscept}). More specifically, the Honda-Owen method extrapolates the measured susceptibility $M/H = \chi_{\text corr} + C_{\text{sat}}M_{\text{sat}}/H$ for $1/H \rightarrow 0$, where $M/H$ is the measured susceptibility, $\chi_{\text corr}$ the corrected, intrinsic susceptibility, $C_{\text{sat}}$ the presumed ferromagnetic impurity content and $M_{\text{sat}}$ its saturation magnetization (see App.~\ref{app:suscept}). Figure~\ref{Chi_corr} shows the temperature dependence of $\chi_{\text corr}$, which is drastically different from $\chi(T)$ in Fig.~\ref{strongfields}a. Specifically, $\chi_{\text corr}(T)$ exhibits a minimum around 65 K and increases almost linearly towards room temperature, signalling a striking non-Curie-Weiss behavior. We note that this increase is qualitatively consistent with the temperature dependence of the magnetic Knight shift in a recent $^{139}$La NMR study on polycrystalline LaNiO$_2$ \cite{Chen2021}, confirming that $\chi_{\text corr}(T)$ is not dominated by a spurious signal of Ni-related impurities. The monotonous increase of $\chi_{\text corr}(T)$ continues at least up to 600 K [Fig.~\ref{FigD2}]. Remarkably, such behavior of $\chi_{\text corr}(T)$ closely resembles the susceptibility of doped cuprates, such as La$_{2-x}$Sr$_x$CuO$_4$ with $x \geq 0.04$ \cite{Nakano1994}. Furthermore, a monotonous increase of the susceptibility was also reported for iron pnictide superconductors and attributed to local AFM correlations \cite{klingeler2010}. In contrast to this monotonously increasing susceptibility, previous studies on polycrystalline LaNiO$_{2}$ \cite{Hayward1999} reported paramagnetic Curie-Weiss behavior between 6 and $300$\,K, including a crossover between two Curie-Weiss regimes at $150$\,K. This discrepancy could be due to a different amounts of impurity phases in the sample of Ref.~\cite{Hayward1999}. In particular, we determine a shorter $c$-axis lattice parameter for our LaNiO$_{2}$ phase [Table \ref{refinement}], suggesting the presence of less excess apical oxygen, possibly due to our longer reduction times (see Methods) and/or the use of CaH$_2$ instead of NaH as a reducing agent. Nevertheless, our extracted absolute value of $\chi_{\text corr}$ at 305\,K of $\approx 4.65\cdot10^{-4}$ emu Oe$^{-1}$ mol$^{-1}$ is comparable to that of Ref.~\cite{Hayward1999}, and also similar to the susceptibility of perovskite LaNiO$_{3}$ \cite{Zhou2014}. Furthermore, $\chi_{\text corr}(T)$ exhibits an upturn below 65\,K [Fig.~\ref{Chi_corr}], which was similarly reported in Ref.~\cite{Hayward1999} and is likely related to paramagnetic impurities.

\subsection{Theoretical modelling}

The observed non Curie-Weiss behavior hints towards the relevance of local and non-local spin-fluctuations as a possible origin of the unusual magnetic response of LaNiO$_2$. In the following, we therefore go beyond effective single-Slater-determinant theories, such as the DFT+$U$ approaches of previous studies \cite{Anisimov1999,Lee2004,Hepting2020,Kapeghian2020,Zhang20202,Botana2020, Liu2020, Ryee2020, Zhang2021}, and turn to quantum-field theoretical methods tailored at the systematic inclusion of temporal and spatial correlations. From their application to the two-dimensional single-band Hubbard model described in Sec.~\ref{sec:theory} we obtain the uniform dc-susceptibility $\chi_{\text{theory}}$ in zero field (see App. \ref{app:comp}). Figure~\ref{fig:theory}a summarizes the results of our calculations as a function of temperature $T$. Initially, we apply DMFT which exhibits (anti)ferromagnetic ordering at $T_{\text{N{\'e}el}}^{\text{DMFT}}\!\approx\!1500$ K as a result of its mean-field nature (see also \cite{Gu20201,Kang2021} for additional DMFT calculations). Above this temperature the uniform susceptibility $\chi^{\text{DMFT}}$ (open blue squares) can be accurately described by a Curie-Weiss law as can be inferred from Fig.~\ref{fig:theory}b, which shows the inverse susceptibility $1/\chi_{\text{theory}}(T)$ and a respective Curie-Weiss fit (dotted black line). The progressive inclusion of spatial correlations on different length scales, however, invalidates this picture: already short-ranged spatial correlations, taken into account by CDMFT on a $4\!\times\!4$ cluster (filled orange circles), lead to a deviation from Curie-Weiss behavior for  $T\!\lessapprox\!T_{\text{N{\'e}el}}^{\text{DMFT}}$. These short-ranged spatial correlations also lead to a small reduction of the onset temperature of magnetic ordering w.r.t. its DMFT value, below which the calculation of $\chi^{\text{CDMFT}}$ is hindered (see App.~\ref{app:comp} for the applied algorithm).

In order to treat correlations larger than the cluster size and eventually enter the low temperature regime, we turn to D$\Gamma$A, which is a diagrammatic extension of DMFT tailored for the systematic inclusion of spatial correlations on all length scales. The susceptibility calculated in D$\Gamma$A (filled red squares)  qualitatively resembles the experimentally determined susceptibility $\chi_{\text corr}(T)$ of LaNiO$_{2}$ [Fig.~\ref{Chi_corr}] with a broad maximum centered around $T_{*}^{\text{D}\Gamma\text{A}}\!\approx\!900$ K, emphasizing  the onset of a strong deviation from Curie-Weiss behavior below that temperature. Note that the maximum of $\chi^{\text{D}\Gamma\text{A}}$ appears {\it without} the emergence of long-range magnetic order, which is prohibited at finite temperatures in 2D systems due to the Mermin-Wagner theorem \cite{Mermin1966,Hohenberg1967}. Nevertheless, strong non-local magnetic fluctuations (paramagnons) can exist in the paramagnetic phase at finite temperatures as a precursor of a $T_{\text{N{\'e}el}}\!=\!0$ transition. These paramagnon contributions are quite naturally included in the D$\Gamma$A framework by its consideration of  ladder-diagrams in both particle-hole channels \cite{Toschi2007}.

Comparing our calculations with the experimentally determined $\chi_{\text corr}$ of LaNiO$_{2}$ presented in Fig.~\ref{Chi_corr}, we find a satisfactory qualitative agreement: both, the rising edge of the susceptibility's broad maximum as well as its relative change in amplitude (between the value at the maximum and at $65$\,K) lie approximately in the same range. This is remarkably accurate given the fact that $\chi^{\text{D}\Gamma\text{A}}$  has been obtained from an effective single-band description. In fact, the same model has been used in the context of superconductivity in NdNiO$_{2}$ \cite{Kitatani2020}, thus corroborating the applicability of a single-band model to IL nickelate compounds. However, a  quantitative comparison of the susceptibilities suggests that overall the experimental response is substantially larger than the calculated one, which will be discussed in more detail below.

\section{Discussion}

The spin glass properties of polycrystalline LaNiO$_{2}$ were revealed by \textmu SR and magnetic susceptibility measurements in small external fields. Whether this glassy behavior is intrinsic to IL nickelates, or induced by impurities and/or secondary phases is an intricate issue. In our PXRD characterization, the signal of possible impurities and secondary phases---including elemental Ni and LaNiO$_{2.5}$---was below the detection threshold. Nevertheless, our magnetization measurements indicated the presence of a small amount of Ni impurities. In principle, Ni or NiO/Ni nanoparticles can show spin glass behavior \cite{Tiwari2005, Aragon2012, Ji2015}, which could rationalize the observed characteristics in the static and dynamic susceptibility in Fig.~\ref{smallfields}. However, the observation of a ZF \textmu SR oscillation with a large relaxation rate $\lambda_{T}$ of up to 25  $\mu s^{-1}$ together with the typical separation of the spectra into a 1/3 and a 2/3 component indicate that the glassy magnetic behavior originates from the bulk of the sample, although with an inhomogeneous spatial distribution. Specifically, the \textmu SR data seem to be incompatible with the scenario that a dilute distribution of Ni nanoparticles induces the observed glassy properties. Moreover, Refs.~\cite{lin202101, Huangfu2020} investigated similar polycrystalline $R$NiO$_2$ and Pr$_4$Ni$_3$O$_8$ powders upon reoxidization and decomposition, concluding that Ni/NiO impurities cannot explain the observed glassy properties. 

Nonetheless, it seems unlikely that square-planar nickelates precisely realize the frustrated nearest and next-nearest neighbor exchange that is required for an intrinsic square lattice spin glass system. Instead, we consider it plausible that local oxygen disorder in form of remaining apical oxygen is responsible for the observed glassy properties. Such excess oxygen can increase the Ni valence towards 2+, yielding an effective moment and consequently a local magnetic cluster, possibly even polarizing the environment to some extent. Furthermore, also entities with a larger spatial extension could be a candidate for the magnetic clusters, including domain walls between regions with differently oriented NiO$_2$ planes or minute inclusions of LaNiO$_{2.5}$. While further work is required to decisively clarify the microscopic origin of the glass state in LaNiO$_{2}$, it is noteworthy that magnetic clusters of varying sizes were also reported in polycrystalline LaNiO$_{2.75 \pm \delta}$ \cite{Sanchez1996,Okajima1995}, which could be a precursor of the clusters observed in LaNiO$_{2}$.  

Along these lines, it is insightful to compare the glassy phase of LaNiO$_{2}$ to the spin glass regime of cuprates. For instance in La$_{2-x}$Sr$_x$CuO$_4$, a glassy phase emerges below a freezing temperature $T_{\text{f}}$ around $5 - 6$\,K for hole doping concentrations $0.02\leq x\leq 0.05$, \textit{i.e.} between the AFM phase and the onset of the superconducting dome  \cite{Harshman1988,Sternlieb1990,Chou1995}. Magnetic resonance experiments attributed the spin glass phase in cuprates to frozen AFM clusters that arise due to the separation of charges into hole-rich and -poor regions \cite{Julien1999}. Thus, the nature of the spin glass in IL nickelates seems to be distinct from the one in cuprates. However, we note that our present study does not allow us to fully rule out a charge segregation scenario in IL nickelates and supplemental measurements on LaNiO$_{2}$ are desirable, for instance with magnetic resonance techniques. 

The magnetic susceptibility measured at high temperatures and in strong magnetic fields provided access to the intrinsic magnetic correlations of LaNiO$_{2}$ [Fig.~\ref{Chi_corr}]. As key experimental result we found that the corrected magnetic susceptibility $\chi_{\text corr}(T)$ follows non-Curie-Weiss behavior and increases above 65 K. Importantly, in absence of long-range order \cite{Hayward1999}, such an increase can be indicative of strong non-local spin fluctuations and is reminiscent of the susceptibility of doped cuprates \cite{Nakano1994} and iron pnictide superconductors \cite{klingeler2010}. It will be interesting to test in future studies whether hole-doping of LaNiO$_{2}$ alters the shape of $\chi_{\text corr}(T)$, similarly to cuprates where the broad maximum in the susceptibility shifts to lower temperatures with increasing doping concentration $x$ and vanishes for heavy doping \cite{Nakano1994}. Moreover, the residual entropy detected in the specific heat $C_{p}$ of LaNiO$_{2}$ [Fig.~\ref{Cpheat}] is compatible with the presence of paramagnons, which are a hallmark feature of doped cuprates. This suggests a possible analogy between the parent (undoped) IL nickelates and doped cuprates.

A good starting point to understand ``non-local spin-fluctuations" from a wave function perspective can be models in the localized limit, such as the Heisenberg model or finite-size clusters. As shown in Fig.~\ref{fig:theory} already a simple $2\!\times\!2$ cluster (and, even more accurately, an embedded cluster via CDMFT) can give rise to broad maxima and deviations from Curie-Weiss behaviour in their static susceptibility without symmetry breaking. This is due to the formation of inter-site singlet ground states and thermal occupation of spin triplets. Another well-known singlet mechanism that leads to a maximum in the static susceptibility is found in the Kondo model where a fully localized magnetic impurity hybridizes with delocalized conduction electrons \cite{hewson_1993}. The ansatz to interpret the $3d$ electrons of Ni as fully localized and oxygen $2p$ or rare-earth $5d$ states as a delocalized bath was addressed in Refs.~\cite{Hepting2020,Zhang20201}. However, when sizable $3d-3d$ hoppings are included in realistic \emph{ab initio} material models, rather a Hubbard-like scenario is supported. In either case the mentioned singlets are non-local non-single-Slater-determinant states. In order to capture their impact on the susceptibility we can therefore neither resort to effective single-particle (like DFT), nor to purely local theories (like DMFT). On the other hand, D$\Gamma$A not only includes short-range spin-fluctuations, such as the mentioned singlets, but fluctuations on all length scales.

Importantly, we find that our D$\Gamma$A calculations [Fig.~\ref{fig:theory}] qualitatively reproduce the salient temperature dependence of the experimental $\chi_{\text corr}$ in Fig.~\ref{Chi_corr}. Therewith, an interpretation of the peculiar susceptibility LaNiO$_{2}$ is provided, since in the framework of D$\Gamma$A the occurrence of a maximum in $\chi^{\text{D}\Gamma\text{A}}$ and a substantial downturn at low temperatures can be attributed to an increasing influence of non-local fluctuations. Furthermore, this lends support to the notion that strong magnetic fluctuations suppress the ordering tendencies in parent IL nickelates and prevent the formation of long-range magnetic order down to lowest temperatures. At a more fundamental level, long-range order is not occurring in our calculations and not accompanying any strong fluctuations at finite temperatures due to the fact that D$\Gamma$A in two dimensions obeys the Mermin-Wagner theorem. Note that this is in contrast to DMFT calculations, where finite temperature ordering is possible due to its mean-field nature with an ordering temperature of $T_{\text{N{\'e}el}}^{\text{DMFT}}\!\approx\!1500 $K. In cuprates, long-range AFM order emerges in spite of a quasi-2D character of the lattice and electronic structure due to anisotropic exchange generating an Ising-like component of the order parameter in compounds with frustrated interlayer exchange \cite{Keimer1992}, or unfrustrated 3D coupling \cite{Tranquada1989}. The absence of apical oxygen and vanishing hybridization between Ni 3$d$ and O 2$p$ states \cite{Hepting2020} might hamper analogous couplings in IL nickelates. The introduction of a hypothetical hopping in the $c$-direction in IL nickelates would lift the 2D constraint of the Mermin-Wagner theorem also for D$\Gamma$A and we would expect to obtain a long-range ordered ground state. However the long-range order would set in at lower temperatures w.r.t. DMFT due to the additional consideration of non-local fluctuations \cite{Rohringer2011, Schaefer2017}.

A comparison of Fig.~\ref{fig:theory} and Fig.~\ref{Chi_corr} shows that the broad maximum in $\chi^{\text{D}\Gamma\text{A}}$ around $T^{\text{D}\Gamma\text{A}}_{*}\!\approx\!900$ K is compatible with the increase of the experimental $\chi_{\text corr}$ with an onset above 65 K. This strengthens the conjecture that the materials class of IL nickelates is a close realization of the single-band Hubbard model \cite{Kitatani2020}. Physically, this remarkable degree of applicability might be rooted in the absence of Zhang-Rice singlets \cite{Hepting2020}, due to the  exceptionally large energetic distance of the oxygen degrees of freedom in IL nickelates \cite{Lee2004,Hepting2020}. However, as already stated in the results section, the magnitude of the theoretically calculated susceptibility is significantly smaller than the experimental determined values. Notably, the different employed computation methods all yield susceptibilities of the same order of magnitude. Thus, the discrepancy to experiment might be explained by our initial assumption for the model, \textit{i.e.} a half-filled Hubbard model with a pure Ni 3$d^9$ configuration. Additional D$\Gamma$A studies exploring multiband effects \cite{Galler2017} and a variation of $U$ are highly desired for addressing this issue.
Furthermore, it cannot be ruled out that $\chi_{\text corr}$ contains contributions beyond such obvious considerations. For instance, it was proposed that topotactically reduced IL nickelates are prone to the inclusion of hydrogen \cite{Si2020}. The resulting high-spin Ni 3$d^8$ contributions --- not captured by our model --- could  enhance the magnetic response substantially. Similarly, the aforementioned remaining apical oxygen due to incomplete reduction can induce local 3$d^8$-like configurations. It is possible that these isolated Ni$^{2+}$ spins are exchange-coupled to the square-planar Ni$^{1+}$ network and therefore exhibit the same temperature-dependent susceptibility, albeit with a larger amplitude.

Along the lines of a possible analogy between undoped IL nickelates and doped cuprates, an evident question is whether the former material class exhibits a  pseudogap similar to that of lightly doped cuprates. More specifically, in cuprates, the hallmark for the original identification of the pseudogap was the suppression of the nuclear magnetic resonance (NMR) Knight shift \cite{Alloul89}. In linear response calculations this suppression corresponds to the formation of a maximum in the uniform magnetic susceptibility \cite{Chen17,Qin2021}. Hence, our complementary CDMFT and D$\Gamma$A analysis indicates that a pseudogap-like regime emerges below $T_{*}^{\text{D}\Gamma\text{A}}\!\approx\!900$ K and the observed non-Curie-Weiss behavior of $\chi_{\text corr}(T)$ is compatible with this notion. Notably, a pseudogap temperature $T_{*}$ higher than 400 K is also consistent with a recent $^{139}$La NMR study on polycrystalline LaNiO$_2$ \cite{Chen2021}. Assuming that our modelization of LaNiO$_{2}$ as a single-band Hubbard model is appropriate, the opening of this pseudogap can be attributed to non-local magnetic correlations \cite{Gunnarsson2015,Schaefer2020,Qin2021}. 

Remarkably, recent RIXS experiments determined the AFM exchange coupling $J$ to be as large as $\sim 65$\,meV in NdNiO$_2$ \cite{lu2021}, which is in good agreement with $77$\,meV obtained from many-body quantum chemistry calculations \cite{Katukuri2020} and reminiscent of the strong exchange of up to $\sim180$\,meV in IL cuprates \cite{Peng2017}. In a simplified picture, considering the strong coupling limit, our choice of parameters corresponds to an exchange coupling of $J = 4t^2/U \sim 198$\,meV. However, such mapping of our model the localized limit of the Heisenberg model and strong coupling limit is likely not appropriate and rather suggests that the coupling is substantially reduced.

\section{Conclusion}

In summary, we presented a comprehensive study of the magnetic correlations in LaNiO$_2$ combining experiment and quantum many-body theories. The observed signatures of strong non-local spin fluctuations, paramagnons, and a pseudogap suggest a striking analogy between the parent IL nickelates and doped cuprates. 

Taking these parallels one step further, the question arises, whether the long-range AFM phase of parent cuprates can be invoked also in IL nickelates, for instance by electron-doping or tuning of the effective bandwidth as well as the Ni-O hybridization. More generally, future multi-method studies using the complementary techniques DMFT, CDMFT and D$\Gamma$A can explore how charge carrier doping affects the magnetic correlations present in parent IL nickelates and in particular to what degree they persist at doping levels that host a superconducting ground state. Moreover, an investigation of the rare-earth series $R$NiO$_2$ can provide insights on whether  hybridization with the rare-earth 5$d$ bands leads to a quantitative deviation from the qualitative applicability of the single-band Hubbard model, which we demonstrated for LaNiO$_2$.

Finally, our study lends support to the notion that AFM spin fluctuations are a prime candidate for mediating superconductivity in IL nickelates.

\section*{Acknowledgements}

We thank R. Merkle and A. Fuchs for the synthesis of LaNiO$_{3}$ and H. Hoier for preliminary PXRD characterizations. We acknowledge PXRD measurements of LaNiO$_{3}$ and LaNiO$_{2}$ by C. Stefani from the X-ray Diffraction Scientific Facility at an early stage of this work. We thank S. Hammoud for ICP-MS measurements and K.~Held, M.~Kitatani, L.~Si, N.~Wentzell and P.~Worm for insightful discussions. In addition, we thank M.~Kitatani for making comparison data available, K.~Held for his inspiring talk on the topic at CNQM2021 and A.~Toschi for intense discussions in Langenlois. We thank the computing service facility of the MPI-FKF for their support and we gratefully acknowledge use of the computational resources of the Max Planck Computing and Data Facility. We gratefully acknowledge the financial support of F. H. by the Swiss National Science Foundation (SNF-Grants No. 200021L-192109). We acknowledge financial support by the Center for Integrated Quantum Science and Technology (IQ$^{\rm ST}$) and the Deutsche Forschungsgemeinschaft (DFG, German Research Foundation): Projektnummer 107745057 - TRR 80.

$^*$ R.A.O. and P.P. contributed equally to this work.

$^{\dagger}$ \href{mailto:Hepting@fkf.mpg.de}{Hepting@fkf.mpg.de} 

$^{\dagger}$ \href{mailto:T.Schaefer@fkf.mpg.de}{T.Schaefer@fkf.mpg.de}

\appendix
\counterwithin{figure}{section}
\section*{Appendix}
\section{Crystal Structure and stoichiometry}
\label{app:structure}

\begin{figure}[t]
\includegraphics[width=.9\columnwidth]{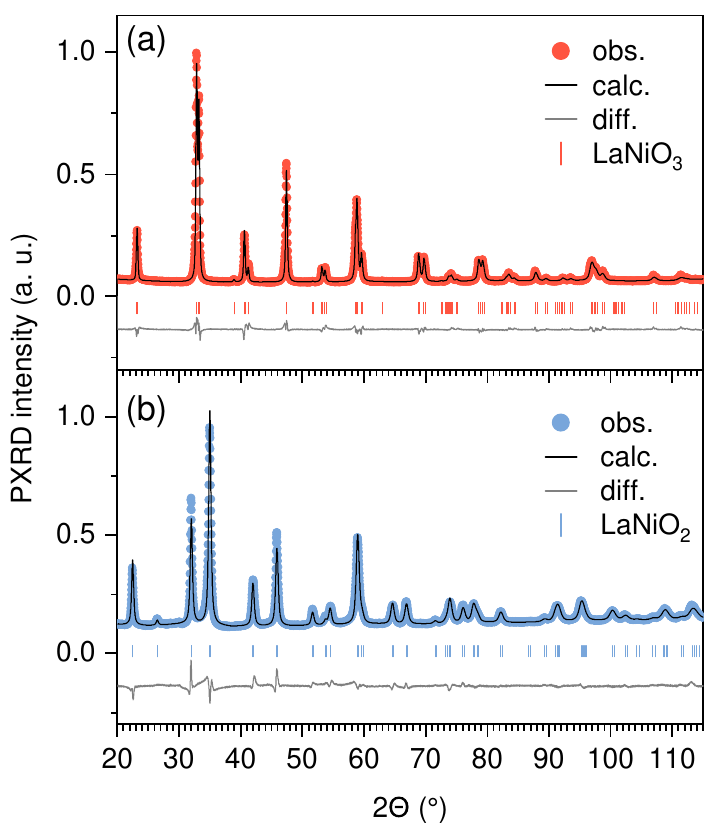}
\caption{Powder X-ray diffraction pattern of LaNiO$_3$ (a) and LaNiO$_{2}$ (b) at $T = 300$ K measured with Cu K$_\alpha$ radiation. Solid black lines correspond to the calculated intensities from the Rietveld refinements. The calculated Bragg peak positions are indicated by vertical bars and the differences between the experimental and calculated intensities are shown as solid gray lines at the bottom.  
}
\label{FigA1}
\end{figure}

\begin{table}
\caption{\label{refinement} (a) Refined atomic coordinates of LaNiO$_{3}$ in rhombohedral space group $R\bar3 c$ (hexagonal axes) extracted from powder X-ray diffraction measurements [Fig.~\ref{FigA1}]. (b) Refined atomic coordinates of LaNiO$_{2}$ in tetragonal space group $P4/mmm$.}

\resizebox{\columnwidth}{!}{\begin{tabular}{ccccc}
\hline
\hline
\multicolumn{5}{c}{(a) LaNiO$_{3}$ $\vert$ $a, b = 5.45095(4)$ $\text{\AA}$, $c = 13.13313(11)$ $\text{\AA}$} \\ \hline
Atom & $x$ & $y$ & $z$ & $U \left( \text{\AA}^2 \right)$ \\ \hline
La (6a) & 0 & 0 & 0.25 & 1.097(8) \\
Ni (6b) & 0 & 0 & 0 & 0.674(17) \\
O (18e) & 0.54430(62) & 0 & 0.25 & 1.637(64) \\
\multicolumn{5}{c}{ }  \\
Reliability factors & $\chi^2$ & $R_B$ & $R_f$ & \\ \hline
 & 7.53(2) & 6.67 &  6.22 & \\ 
   &  &  &  &  \\
\hline
\hline
\multicolumn{5}{c}{(b) LaNiO$_{2}$ $\vert$ $a, b = 3.95496(4)$ $\text{\AA}$, $c = 3.36582(5)$ $\text{\AA}$}\\ \hline
Atom & $x$ & $y$ & $z$ & $U \left( \text{\AA}^2 \right)$ \\ \hline
La (1d) & 0.5 & 0.5 & 0.5 & 0.644(10) \\
Ni (1a) & 0 & 0 & 0 & 0.494(31) \\
O (2f) & 0 & 0.5 & 0 & 0.039(67) \\
\multicolumn{5}{c}{}   \\
Reliability factors & $\chi^2$ & $R_B$ & $R_f$ & \\ \hline
 & 6.61(2) & 7.34 &  5.06 & \\ 
\end{tabular}}
\end{table}

Powder X-ray diffraction (PXRD) of LaNiO$_{3}$ and the reduced sample (LaNiO$_{2}$) was performed at room temperature using a Rigaku Miniflex diffractometer with Cu K$_\alpha$ radiation [Fig.~\ref{FigA1}]. Rietveld refinements were conducted with the FullProf software suite \cite{Rodriguez1993}. LaNiO$_{3}$ and LaNiO$_{2}$ were refined in rhombohedral space group $R\bar3c$ and tetragonal $P4/mmm$, respectively. The refined structural parameters are presented in Table \ref{refinement}. PXRD did not indicate the presence of any impurity phases.

\begin{figure}[t]
\includegraphics[width=.9\columnwidth]{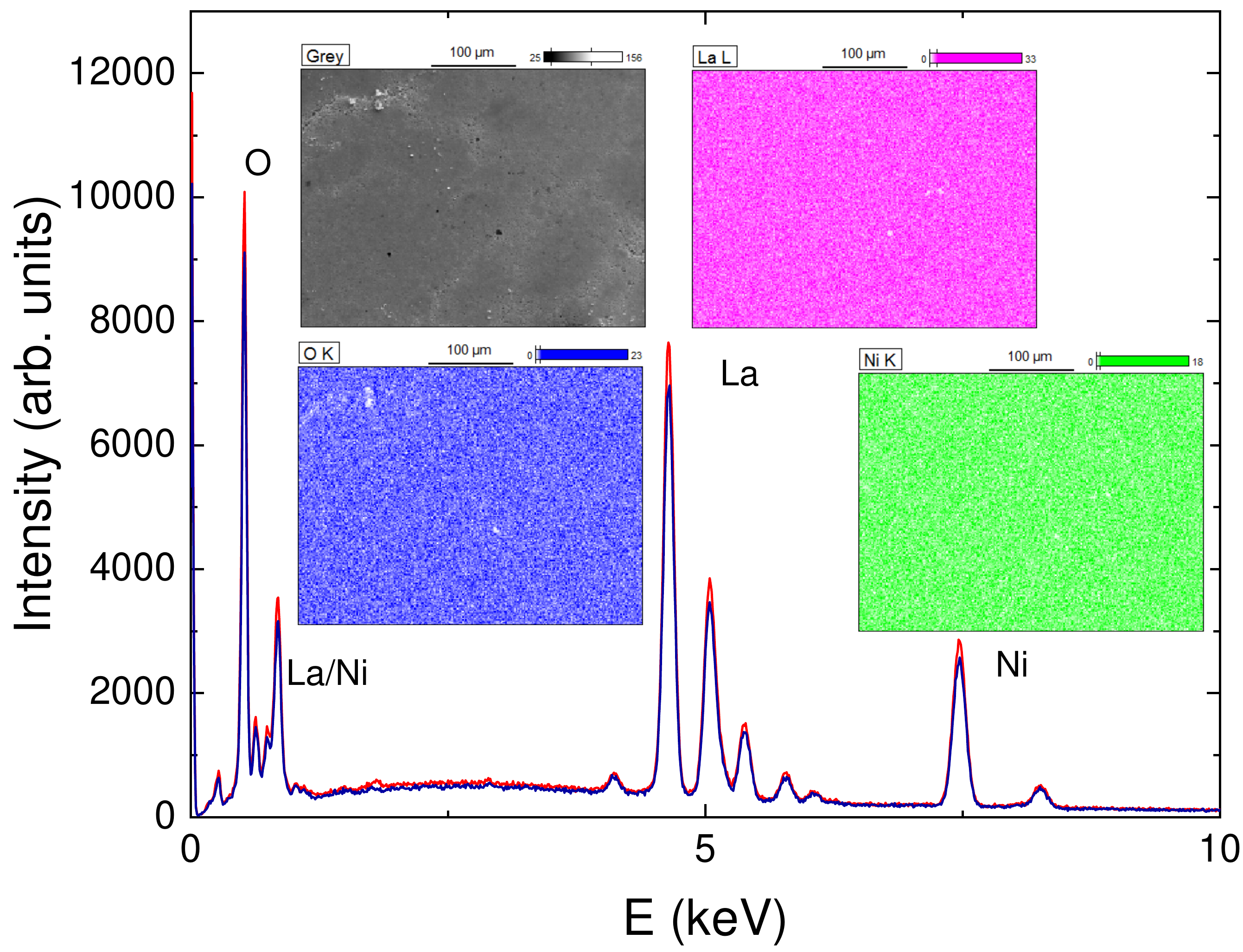}
\caption{Energy dispersive x-ray spectra (EDS) of a pellet of LaNiO$_2$. The red line corresponds to an area scan, while the blue line is from a local point. The insets show the EDS maps of different elements, with Ni in green, La in magenta, and O in blue. No sudden jumps in Ni compared to La were observed.  
}
\label{FigA2}
\end{figure}

Moreover, the stoichiometry of the samples was determined using scanning electron microscopy (SEM) with energy dispersive x-ray spectroscopy (EDS) [Fig.~\ref{FigA2}] and inductively coupled plasma mass spectroscopy (ICP-MS). The latter method indicates the stoichiometries La$_{0.99(1)}$Ni$_{0.99(1)}$O$_{2.99(3)}$ and La$_{0.99(1)}$Ni$_{0.99(1)}$O$_{2.02(3)}$, corresponding to the ideal stoichiometries (within the experimental error) of the perovskite and infinite-layer phase, respectively. The former method indicates that a pressed LaNiO$_{2}$ pellet exhibits a highly homogeneous distribution of all elements down to the resolution limit [see inset in Fig.~\ref{FigA2}]. Together with PXRD, we can estimate that ferromagnetic Ni inclusions in the LaNiO$_{2}$ sample are below 1 wt\%.

\section{Complementary \textmu SR data}
\label{app:usr}

\begin{figure}[t!]
\centering
\includegraphics[width=.75\columnwidth]{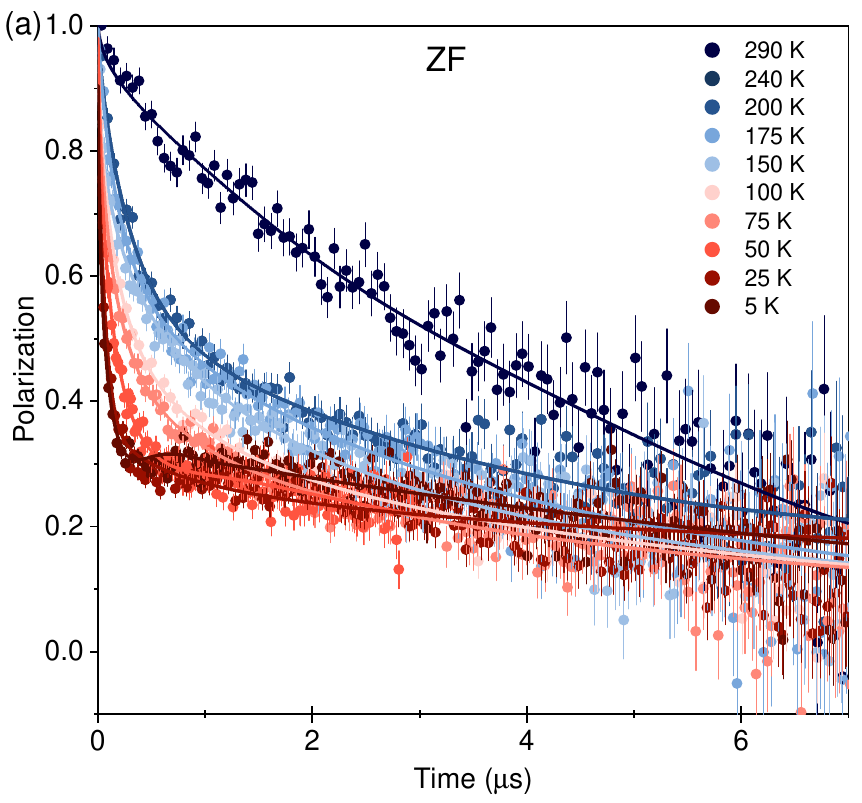}
\caption{ZF \textmu SR spectra measured at various temperatures. Solid lines are fits to the data (see text). }
\label{FigU2}
\end{figure}

Figure~\ref{FigU2} shows the full set of ZF \textmu SR spectra measured at various temperatures.

\begin{figure}[t!]
\centering
\includegraphics[width=.75\columnwidth]{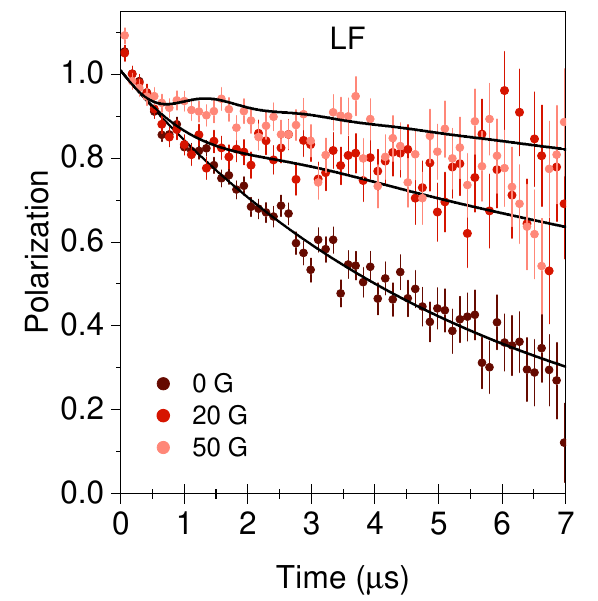}
\caption{\textmu SR spectra at 290 K in different longitudinal fields (LF). Solid black lines are dynamical fits (see text). }
\label{FigU1}
\end{figure}

Figure~\ref{FigU1} displays \textmu SR spectra measured at 290 K in different longitudinal fields (LF). The indicated fits with a dynamical Lorentzian Kubo-Toyabe model describe the data reasonably well, whereas a static model is inconsistent with the data (not shown here). From the fit we obtain a fluctuation rate of approximately 0.5 MHz at 290 K.

\section{Complementary susceptibility data}
\label{app:suscept}

\begin{figure}[t!]
\centering
\includegraphics[width=1.0\columnwidth]{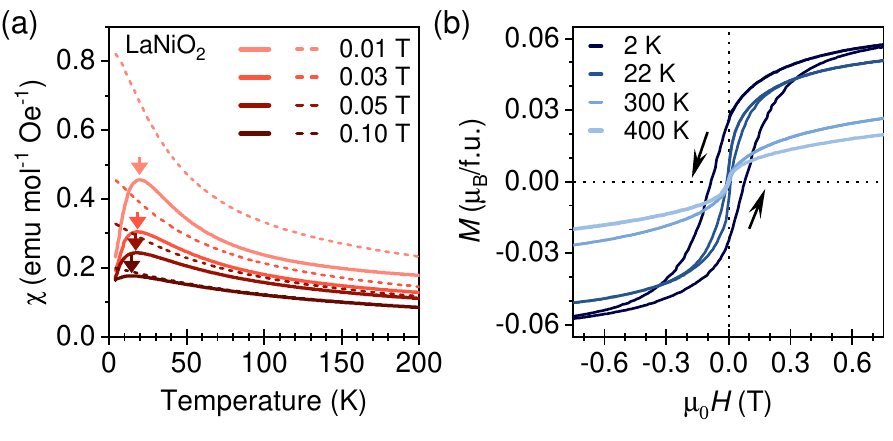}
\caption{(a) Static (dc) magnetic susceptibility $\chi$ of LaNiO$_{2}$ in different external magnetic fields measured upon heating after zero-field cooling (ZFC, solid lines) and field-cooling (FC, dashed lines), respectively. The arrows indicate the positions of the cusps in the ZFC curves. (b) Isothermal magnetic hysteresis loops of LaNiO$_{2}$ measured up to $\mu_0 H = 0.75$ T at different temperatures. Black arrows indicate the directions of the field sweep.}
\label{FigB1}
\end{figure}

Figure~\ref{FigB1}a shows the temperature dependence of the susceptibility $\chi$ of LaNiO$_{2}$ in different external magnetic fields. As expected for a spin glass, upon increasing field, the cusp in the ZFC curves shifts to lower temperatures and the ZFC and FC curves join together at lower temperatures. 

We identify two different regimes where the irreversibility of the ZFC and FC curves is present [Fig.~\ref{smallfields}a]. In the low temperature regime around $T_{\text{f}}$ and below, the width of the hysteresis in the isothermal magnetic hysteresis loops is particularly broad [Fig.~\ref{FigB1}b] and we attribute the ZFC - FC splitting mostly to the glassy character of the system. In general, spin glasses can exhibit a remnant magnetization in the temperature region of the freezing transition and below, due to the irreversible behavior of the magnetization. This remnant magnetization depends in a detailed way on the ``magnetic history" of the sample \cite{Huangfu2020} and evolves with time; \textit{i.e.} it does not reflect the thermal equilibrium behavior of the glass. Such hysteretic effects can occur not only in spin glasses with underlying ferromagnetic correlations, but also in the antiferromagnetic case, thus precluding such distinction in our measurement. In antiferromagnetic glasses, hysteresis can be observed in particular in case of slow dynamics and/or uncompensated spins at surfaces of clusters.

The persisting ZFC-FC splitting in the regime above $T_{\text{f}}$ in Fig.~\ref{FigB1}a is possibly induced by small amounts of ferromagnetic impurities, such as elemental Ni. This splitting is relatively narrow and also the widths of the hysteresis loops decrease significantly above $T_{\text{f}}$ [Fig.~\ref{FigB1}b]. Nevertheless, it is also possible that the observed ZFC-FC splittings and hysteresis do not correspond to two different regimes, but are a consequence of either, the spin glass or Ni impurities. Moreover, a coupling between the two phenomena could be responsible for the enhanced width of the hysteresis at low temperatures. 

Note that our sample characterization (see Methods and App.~\ref{app:structure}) indicates that the amount of (ferromagnetic) impurities is very small. This notion is supported by the magnetization measurements shown in Fig.~\ref{FigB1}b, revealing that the saturation magnetization in units of $\mu_{B}$/Ni is low. In particular, the highest observed magnetization values are still substantially smaller than the expected value of $0.6\mu_{B}$ for metallic Ni, ruling out the presence of a bulk ferromagnetic Ni phase. Hence, we conclude that the Ni impurity contribution in our sample is significantly less than 2.5 wt\%.

\section{Corrected susceptibility and model fits}
\label{app:fit}

\begin{figure}[t!]
\centering
\includegraphics[width=.8\columnwidth]{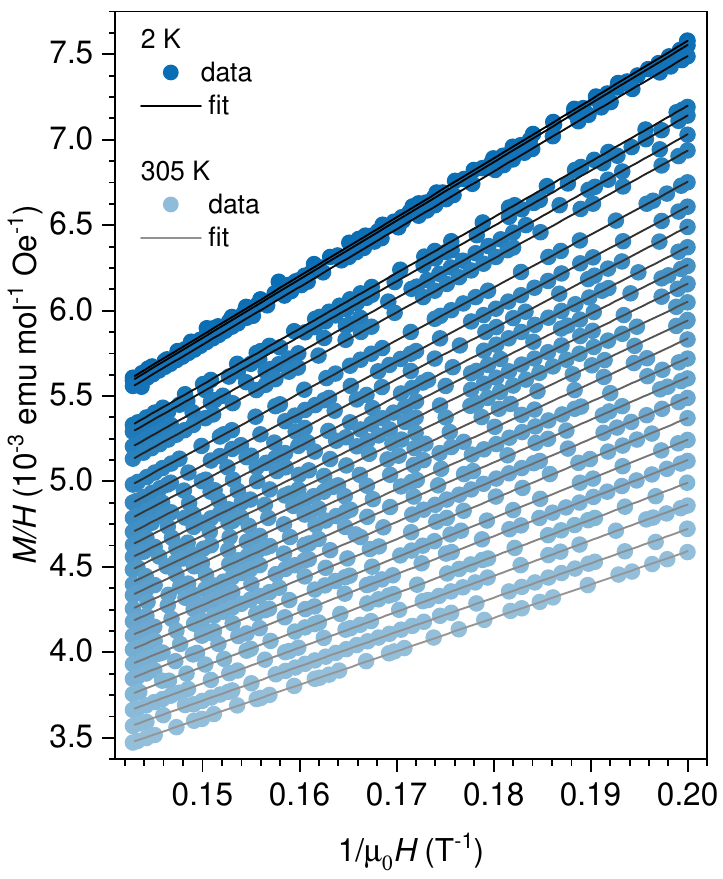}
\caption{Isothermal magnetization of LaNiO$_{2}$ measured between 5\,T and 7\,T at 2\,K, 4\,K, 10\,K, 15\,K, 20\,K, 35\,K, 50\,K, 65\,K, 80\,K, 95\,K, 110\,K, 125\,K, 140\,K, 155\,K, 170\,K, 185\,K, 200\,K, 215\,K, 230\,K, 245\,K, 260\,K, 275\,K, 290\,K, and 305\,K, respectively (filled symbols from dark blue to light blue). The data ($M/H$) are plotted as a function of $1/H$, according to the Honda-Owen method. Solid lines (from dark to light gray) are linear fits for each temperature. The slope of each linear fit corresponds to the corrected paramagnetic susceptibility $\chi_{\text corr}$ displayed in Fig.~\ref{Chi_corr} in the main text.}
\label{FigD1}
\end{figure}

Figure~\ref{FigD1} shows the raw data of the isotherms of LaNiO$_{2}$ together with the fits from which the corrected susceptibility $\chi_{\text corr}(T)$ in Fig.~\ref{Chi_corr} of the main text was extracted. We used the Honda-Owen method to determine the intrinsic susceptibility from the magnetization isotherms measured at high magnetic fields. We consider two contributions in the measured magnetization: one originating from the intrinsic paramagnetic susceptibility and the other from ferromagnetic impurities, corresponding to
\begin{equation}
    M = \chi_{\scriptsize corr}H + C_{\textrm{\scriptsize sat}}M_{\textrm{\scriptsize sat}},
\end{equation}
where $\chi_{corr}$ is the intrinsic susceptibility, $H$ is the applied magnetic field and $M_{\text{sat}}$ is the magnetization due to impurities. Then, in the limit $\mu_0 H \to \infty$, the slope in a $M/H$ vs. $1/\mu_0 H$ plot corresponds to the intrinsic susceptibility $\chi_{corr}$ (See Fig.~\ref{FigD1}). We note that in Ref.~\cite{Hayward1999} a method  (``subtraction method") closely similar to the Honda-Owen method was applied to extract the susceptibility of polycrystalline LaNiO$_{2}$; however, from magnetization isotherms measured only up to 5 T. 

\begin{figure}[t!]
\centering
\includegraphics[width=0.95\columnwidth]{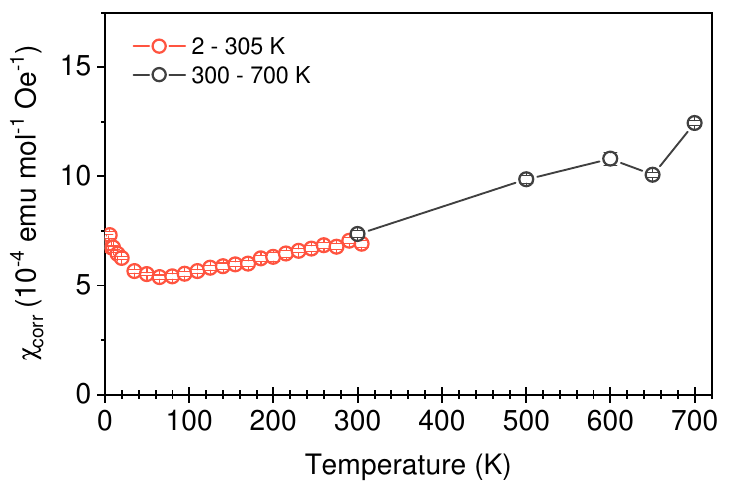}
\caption{Corrected paramagnetic susceptibility $\chi_{\text corr}$ of LaNiO$_{2}$ across a wide temperature range. Red data points are reproduced from Fig.~\ref{Chi_corr} in the main text. Dark gray data points were extracted from isotherms [Fig.~\ref{FigDX}] measured in a high-temperature setup (see Methods). 
}
\label{FigD2}
\end{figure}

\begin{figure}[t!]
\centering
\includegraphics[width=0.90\columnwidth]{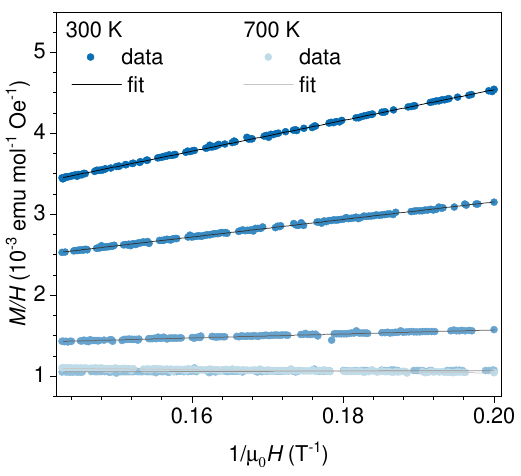}
\caption{Isothermal magnetization measured between 5\,T and 7\,T at 300\,K, 500\,K, 600\,K, 650\,K, and 700\,K, respectively (filled symbols from dark blue to light blue). Solid lines (from dark to light gray) are linear fits for each temperature to extract $\chi_{\text corr}$ (see Fig~\ref{FigD2}). The data ($M/H$) are plotted as a function of $1/\mu_0H$, according to the Honda-Owen method.
}
\label{FigDX}
\end{figure}

In addition to the data shown in Fig.~\ref{FigD1}, isotherms between 5 to 7 T were measured at elevated temperatures and the magnetic susceptibility was extracted. As can be seen in Fig.~\ref{FigD2} and Fig.~\ref{FigDX}, the monotonous increase of $\chi_{\text corr}(T)$ of LaNiO$_{2}$ continues at least up to 600 K. At even higher temperatures, the small dip in the curve around 650 K is likely associated with the presence of Ni impurities, which are estimated to be significantly less than 2.5 wt\% in our pristine LaNiO$_{2}$ sample (see App.~\ref{app:suscept}) while additional impurities might be created during the heating process. These Ni impurities apparently have a small effect on $\chi_{\text corr}(T)$, even though $\chi_{\text corr}$ is extracted via the Honda-Owen method from high-field isotherms between 5 to 7 T. Note that the increase after 650 K in signal is due to the oxidization and decomposition of the sample as revealed by repeated measurement and PXRD analysis.

We conclude that the temperature dependence of $\chi_{\text corr}$ of our LaNiO$_2$ sample well below 600 K is not qualitatively dominated by the Ni impurities. Specifically, we note that the temperature dependence of the magnetic Knight shift in a recent $^{139}$La NMR study on polycrystalline LaNiO$_2$ \cite{Chen2021} is qualitatively consistent with our $\chi_{\text corr}(T)$ in Fig.~\ref{FigD2}, indicating that our $\chi_{\text corr}(T)$ essentially reflects the intrinsic susceptibility of LaNiO$_2$ and not that of Ni impurities.

\section{Computational details of the numerical calculations}
\label{app:comp}

\subsubsection{DMFT and D\texorpdfstring{$\Gamma$}{Gamma}A}
We apply D$\Gamma$A in its ladder-version with Moriyaesque $\lambda$-corrections in the spin channel \cite{Toschi2007, Katanin2009, Rohringer2018b}, since it is particularly suited for the description of low-dimensional systems with strong non-local magnetic fluctuations in the paramagnetic phase (paramagnons) \cite{Toschi2007,Schaefer2015,Schaefer2015b,Schaefer2021} due to its incorporation of non-local ladders in the relevant  particle-hole channels \cite{Toschi2007}. We solve the Bethe-Salpeter equations in Matsubara frequency space with a maximum of $N_{i\omega}\!=\!100$ positive fermionic and $N_{i\Omega}\!=\!99$ positive bosonic Matsubara frequencies for the two-particle Green function at the lowest temperature shown. We extrapolated the physical susceptibility in the number of fermionic Matsubara frequencies $N_{i\omega}\!\rightarrow\!\infty$ according to $\chi\!=\!a+b/N_{i\omega}^2$. We used $128$ linear momentum grid points. In our calculations the dc-susceptibility corresponds to the zeroth bosonic Matsubara frequency and zero momentum transfer, i.e. $\chi^{\text{D}\Gamma\text{A}}\!\equiv\!$ Re $\chi_m(\mathbf{q}\!=\!(0,0),i\Omega_n\!=\!0)$. For the solution of the self-consistently determined Anderson impurity model we used the  continuous-time quantum Monte Carlo solver in its interaction expansion CT-INT \cite{Rubtsov2004,Rubtsov2005,Gull2011a} as part of the TRIQS \cite{TRIQS} package. We used $256 \cdot 10^5$ cycles and $6200$ core hours per temperature point.

\subsubsection{CDMFT}

In order to calculate the magnetic susceptibility in CDMFT, analogously to experiment, we applied a small ferromagnetic (uniform) field with strength $H_F=0.008$ and measured the mean uniform magnetization $m$ of the cluster:

\begin{equation}
    \text{Re }\chi_m(\mathbf{q}\!=\!(0,0),i\Omega_n\!=\!0) = \left.\frac{\partial m}{\partial H}\right\vert_{H=0} \approx \frac{m}{H_F}.
\end{equation}

We checked that this value of the applied field is within the linear response regime. We again used the continuous-time quantum Monte Carlo solver in its interaction expansion CT-INT as part of the TRIQS package. We converged the calculations using $30$ iterations with a statistic of $256 \cdot 10^6$ cycles for each iteration and $4000$ total core hours per temperature point.

\subsubsection{Unit conversion}
We converted the computed susceptibilities given in units of [$\mu_{\text{B}}^2$ eV$^{-1}$] to [emu mol$^{-1}$ Oe$^{-1}$] by multiplying with the factor $3.233\cdot 10^{-5}$ [emu mol$^{-1}$ Oe$^{-1}$ $\mu_{\text{B}}^{-2}$ eV].

\bibliographystyle{apsrev4-2}
\bibliography{Literature}

\end{document}